\documentclass[ba]{imsart}
\pubyear{2022}
\volume{TBA}
\issue{TBA}
\firstpage{1}
\lastpage{1}

\usepackage{amsthm}
\usepackage{amsmath}
\usepackage{natbib}
\usepackage{bbm}
\usepackage[colorlinks,citecolor=blue,urlcolor=blue,filecolor=blue,backref=page]{hyperref}
\usepackage{graphicx}

\usepackage[utf8]{inputenc}
\usepackage{amsmath}
\usepackage{amssymb}
\usepackage[normalem]{ulem}
\usepackage{natbib}
\usepackage{color} 
\usepackage[dvipsnames]{xcolor}
\usepackage{csvsimple}
\usepackage{siunitx,array,booktabs}
\usepackage{numprint}

\usepackage{subcaption}

\usepackage{rotating}
\usepackage{tikz}

\DeclareMathOperator{\s}{\mathbf{s}}
\DeclareMathOperator{\x}{\mathbf{x}}
\DeclareMathOperator{\y}{\mathbf{y}}
\DeclareMathOperator{\z}{\mathbf{z}}

\startlocaldefs
\numberwithin{equation}{section}
\endlocaldefs

\begin{document}


\begin{frontmatter}

\title{Species Distribution Modeling with Expert Elicitation and Bayesian Calibration}
\runtitle{SDMs with Expert Elicitation and Bayesian Calibration}

\begin{aug}
\author{\fnms{Karel} \snm{Kaurila}\thanksref{addr1}},
\author{\fnms{Sanna} \snm{Kuningas}\thanksref{addr2}},
\author{\fnms{Antti} \snm{Lappalainen}\thanksref{addr2}},
\and
\author{\fnms{Jarno} \snm{Vanhatalo}\thanksref{addr1,addr3}}


\runauthor{}

\address[addr1]{Department of Mathematics and Statistics, University of Helsinki}
\address[addr2]{Natural Resources Institute Finland}
\address[addr3]{Organismal and Evolutionary Biology Research Programme, University of Helsinki}


\end{aug}

\begin{abstract}
Species distribution models (SDMs) are key tools in ecology, conservation and management of natural resources. 
They are commonly trained by scientific survey data but, since surveys are expensive, there is a need for complementary sources of information to train them.
To this end, several authors have proposed to use expert elicitation since
local citizen and substance area experts can hold valuable information on species distributions. 
Expert knowledge has been incorporated within SDMs, for example, through informative priors.
However, existing approaches pose challenges related to assessment of the reliability of the experts.
Since expert knowledge is inherently subjective and prone to biases, we should optimally calibrate experts' assessments and make inference on their reliability.
Moreover, demonstrated examples of improved species distribution predictions using expert elicitation compared to using only survey data are few as well.
In this work, we propose a novel approach to use expert knowledge on species distribution within SDMs and demonstrate that it leads to significantly better predictions. 
First, we propose expert elicitation process where experts summarize their belief on a species occurrence proability with maps. 
Second, we collect survey data to calibrate the expert assessments. Third, we propose a hierarchical Bayesian model that combines the two information sources and can be used to make predictions over the study area.
We apply our methods to study the distribution of spring spawning pikeperch larvae in a coastal area of the Gulf of Finland. 
According to our results, the expert information significantly improves species distribution predictions compared to predictions conditioned on survey data only. 
However, experts' reliability also varies considerably, and even generally reliable experts had spatially structured biases in their assessments.
This suggests that expert elicitation can be an efficient tool, for example, in natural resources management and conservation area planning, but expert information should be analyzed with care and preferably calibrated.
\end{abstract}




\end{frontmatter}

\section{Introduction}

Species distribution models (SDMs) are key tools in ecology, conservation and management of natural resources. 
They are used to study, for example, species habitat preferences \citep{Kallasvuo_etal:2017,elith:2009}, interspecific interactions \citep{vanhatalo2020,Ovaskainen+abrego:2020} and to build species distribution maps by predicting species presence and abundance over extended regions where species data have not been collected \citep{Kotta2019,Makinen+Vanhatalo:2018,gelf+sil:2006}. 
SDMs are commonly trained by controlled survey data but, since organizing scientific surveys is expensive and human labor intensive, there is also a need for complementary sources of information on species distributions.
One such source is expert knowledge which has been used to inform species distribution predictions either independently or together with survey data \citep{murray2009useful,pearman2020predicting,Crawford+etal:2020}. However, expert knowledge is inherently subjective posing challenges related to model calibration \citep{murray2009useful,DiFebbraro+etal:2018}. Moreover, it is not clear what type of expert knowledge should be collected and who are the experts to trust \citep{Pearce+etal:2001,DiFebbraro+etal:2018}.

In this work, we propose a novel approach to elicit expert knowledge on species distribution and to calibrate it using survey data within hierarchical Bayesian framework.  
Expert elicitation refers to a process during which a statistician extracts probability distributions, or point estimates, for a parameter or a variable of interest from one or more people who are believed to have valuable information about them. 
Comprehensive, general, treatments of modern approaches to expert elicitation are given by \citet{OHagan+etal:2006} and \citet{Dias_etal:2018}.
Expert elicitation is extensively used in science, management and other applications \citep[e.g.,][]{Burgman+etal:2011,Vanhatalo_etal:2014,Nevalainen_etal:2018,OHagan:2019,perala2020,LaMere_etal:2020a}.
In the context of species distribution modeling, expert elicitation has been used especially in conservation and management applications. For example, \citet{Crawford+etal:2020} used expert opinion to inform habitat suitability models for concervation planning in US and \citet{DiFebbraro+etal:2018} assessed the feasibility of monitoring habitat quality for bird communities in Central Italy by using survey data and expert driven models. \citet{pearman2020predicting} used expert elicitation to model the distribution of several species in New England. They used a web questionnaire to elicit species presence probabilities and information on the impact of covariates on the species occurency. When fitting the SDM, the assessments were pooled together by weighting them based on the experts' self-assessment on their confidence.
\citet{murray2009useful} elicited informative priors from experts for a species distribution model, where the presence or absence of a threatened species was modelled with logistic regression. They provided experts with an interactive tool to tune the parameters of the beta distribution to give their assessment of species presence in various environments.

We will move forward from the above mentioned approaches by explicitly modeling and correcting for the systematic biases in the expert assessments. 
Correcting for possible biases is crucial for reliable use of expert information in inference since humans are prone to psychological idiosyncrasies and subjective biases \citep{Dias_etal:2018,Burgman+etal:2011,Tversky+Kahneman:1974}. 
We will follow the so called \emph{supra Bayesian} approach, where expert information is treated as observations which are linked to model parameters through a conditional probability distribution \citep[likelihood function;][]{Genest+Schervish:1985,Lindley+Singpurwalla:1986,French2011,albert2012}. 
Our approach is based on three components: i) expert elicitation process which summarizes experts' own assessment of their region of expertize and belief on species occurrence proability within that area, ii) data from carefully designed survey within the study region that can be used to calibrate the expert assessments, and iii) a hierarchical Bayesian model that combines the two information sources.

\begin{figure}[t]
         \centering
         \includegraphics[width=\textwidth]{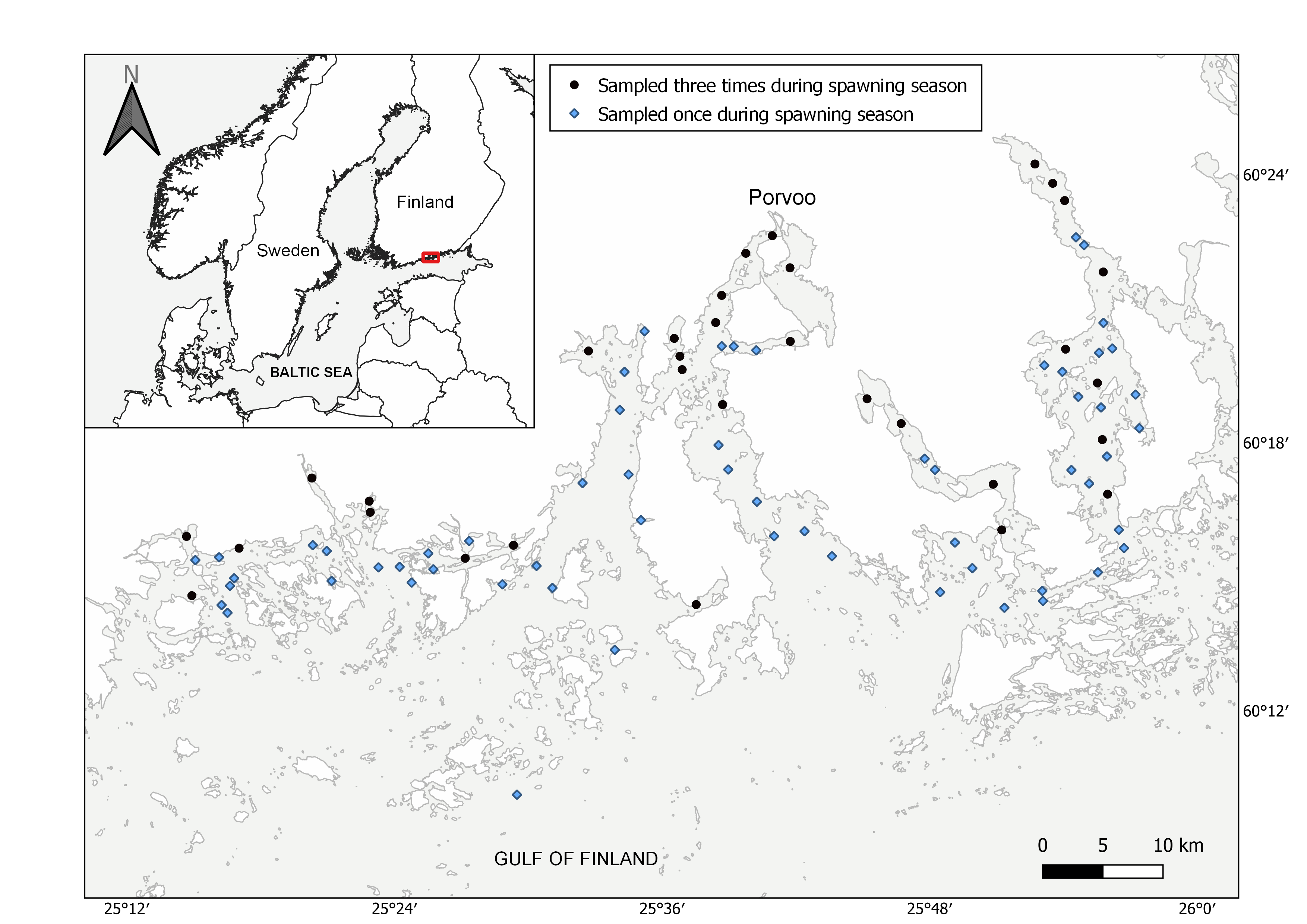}
\caption{The study area in Porvoo-Sipoo archipelago, in the Baltic Sea. The dots show the locations from where the survey data was collected.}\label{fig:study_area}
\end{figure}

We motivate our approach with a real wold case study where we map the  distribution of newly hatched pikeperch (\emph{Sander lucioperca}) larvae within the Porvoo-Sipoo fisheries region (a public corporation whose purpose is to develop fishery in their region) in the Gulf of Finland (see Figure~\ref{fig:study_area}).
Pikeperch is a top predator of Baltic Sea coastal areas and central species in the coastal ecosystem of the Baltic Sea. 
The coastal pikeperch population forms a commercially important fish stock and pikeperch are also highly sought after by recreational fishers. 
Hence, knowledge about its spawning areas (i.e., areas with newly hatched larvae) has both economical and conservation importance. 
Pikeperch is of fresh water origin and has specific habitat requirements for their reproduction, selecting shallow ($<$10 m deep), vegetated, and sheltered bays that warm up early in the spring \citep{Veneranta_etal:2011,Kallasvuo_etal:2017}.
Information on pikeperch spawning areas is important for implementation of conservation measures to protect the spawning, the larvae and male who are securing the development of their offspring.
Maps of the spawning areas can be used to guide local fisheries management and coastal area planning \citep{Kallasvuo_etal:2017} and our aim is to improve these maps by combining survey data with expert information.

The rest of the paper is organized as follows. We first summarize the data collection and expert elicitation process used in our motivating case study. After that we propose a hierarchical Bayesian model for these data and methods to conduct model comparison. We then present the results from our case study, discuss them and provide some concluding remarks.


\section{Data collection and expert elicitation}

\subsection{Survey data and environmental covariates}\label{sec:survey_data}

The study area consists of coastal environment types ranging from open water to sheltered bays. 
In our analysis, we used three environmental covariates to characterize the environment: \emph{depth} (m), \emph{distance to deep} water ($>$10 m) and \emph{shoreline density} (km/m$^2$). The latter two covariates are proxies for how close to shoreline and how sheltered a location is. Each of the covariates were available throughout the study area as raster maps with 50 m resolution. More detailed description of the covariates and how they were constructed is provided by \citet{Kallasvuo_etal:2017}.

To collect survey data on the distribution of the larvae of pikeperch from the study area we conducted a field survey of the surface water layer in June 2017 with paired Gulf ichthyoplankton samplers. 
These are small nets that are pulled by a small boat over a transect of 500 meters and have been used to quantitatively monitor the abundance and spatial occurrence of pikeperch larvae also in earlier studies \citep[see][for a more detailed description]{Veneranta_etal:2011,Kallasvuo_etal:2017,langnabba_etal:2019}. 
We sampled in total 92 sites (Figure~\ref{fig:study_area}) which were dispersed over the entire study area covering all main habitat types. 
The sampling was scheduled to the \emph{a priori} estimated peak larval season \citep{Liu+Vanhatalo:2020} in June. 
From each sample, we counted early-stage larvae (size range of 3–17 mm) and recorded the effort; that is, the volume of water sampled, which equals the length of the transect multiplied by the size (area) of the opening of the Gulf ichthyoplankton sampler.
The Gulf ichthyoplankton samplers are accurate method for larval sampling and the sampling times were set so that the variability in catchability due to weather conditions was minimized \citep{langnabba_etal:2019}. Hence, data collected with the Gulf ichthyoplankton samplers will be treated as the ground truth in this study.

\subsection{Expert elicitation}\label{sec:expert_elicitation}

We elicited information concerning the pikeperch spawning grounds from ten active fishermen (to be called experts hereafter) whose fishing areas fall inside our study area. 
The experts were suggested by the executive director of the local fisheries region. Before the elication, the executive director also confirmed by a phone call to all experts that the experts were well motivated to participate in the elicitation process.
We sent to each expert by post a printed map of the study area, crayons with three different colors and filling instructions.
First the experts were asked to draw the borders of their assessment regions to the map; assessment regions were defined to be areas within which an expert was confident to state his/her beliefs. 
After this, each expert was asked to color the areas, within their assessment region, where they believed pikeperch did or did not spawn. 
The experts were asked to describe the strength of their belief using four categories that were described as follows (translated from the original Finnish and Swedish versions):
\begin{itemize}
 \item(Colour 1) A generally known, locally or regionally important pikeperch spawning area (the probability that pikeperch spawns in the area is over 90\%)
 \item (Colour 2) Another area where pikeperch most likely spawns (the probability that pikeperch spawns in the area is 50 -- 90 \%)
 \item (Colour 3) An area where pikeperch might spawn (the probability that pikeperch spawns in the area is 10 -- 50\%)
 \item (Uncoloured) The areas that are not marked to belong to any of the above three categories, but are inside an assessment region, are considered areas where pikeperch does not spawn according to the general knowledge (the probability that pikeperch spawns in the area is less than 10\%)
\end{itemize}

Sea areas outside an expert's assessment regions were considered as missing information from that expert. 
Figure~A.1 shows an example of an expert elicitation form and a map drawn by an expert. 
The elicitation process and elicitation questions were tested before the actual elicition with the executive director of the local fisheries region. 
We then revised the questions and the elicitation process according to her feedback.
The written descriptions of the colour categories were particularly carefully planned together so that they would be correctly understood by the fishermen.
The expert elicitation was organized in spring 2018 after which we digitized the expert drawn maps by scanning them and storing them as raster maps. We aligned the expert assessment raster maps with the raster maps of the environmental covariates (Section~\ref{sec:survey_data}) using the ArcGIS software so that each expert's answers formed one raster map layer whose lattice match that of the environmental covariates. 
The grid cells of the expert assessment maps were labeled by one of the above four categories or NA (see Figure~A.2). 
The latter denotes an area outside an expert's assessment region. 
There are no significant errors from the digitalization since the questionnaire maps were printed in the same coordinate system as the raster maps of the environmental covariates.

The motivation behind asking directly about spawning areas and the reasoning for using the four above classes to describe experts' knowledge were the following. 
Earlier works have indicated that expert information is most accurate within areas from where experts have personal experience \citep{Pearce+etal:2001,DiFebbraro+etal:2018,Crawford+etal:2020}.
It has also been shown that experts are typically better in assessing their belief on real world variables, which they can observe, than on parameters of statistical models \citep{OHagan+etal:2006}. The latter are mathematical abstractions whose interpretation requires expertise in modeling. 
Hence, we wanted the experts to directly assess their beliefs concerning a variable that has a clear undebatable real world meaning instead of asking them to provide prior information for some parameters of our species distribution model (Section~\ref{sec:SDMs}). 
The definition of pikeperch spawning area is clear and it is understood in uniform manner by both the fishermen and researchers. 
Even though the fishermen cannot directly observe pikeperch spawning they are assumed to be skilled to assess spawning areas based on their personal experience on where pikeperch concentrate during spawning season.
We also wanted to allow experts to express the uncertainty in their knowledge since this provides a more complete picture of their beliefs than hard cut division to spawning and no-spawning areas \citep[see also][]{OHagan+etal:2006}. 
However, since filling in the questionnaire maps was rather tedious we wanted to restrict the level of detail. 
The four categories described above provide a compromise between these two targets.


\section{Species distribution models}\label{sec:SDMs}

\subsection{Model for survey data}

\subsubsection{Model for larval counts}

We followed \citet{Liu+Vanhatalo:2020} and \citet{Kallasvuo_etal:2017} and modeled the distribution of pikeperch larvae with a log Gaussian Cox process (LGCP) with intensity function $\lambda(\s,\x(\s))$ where $\s$ denotes a location inside the study area and $\x(\s)$ is a vector of environmental covariates at that location. 
We modeled the log intensity with a linear function of the covariates and a spatial random effect:
\begin{equation}\label{eq:transect_obs_intensity}
\log \lambda(\textbf{s}_i,\textbf{x}_i) = \alpha+\boldsymbol{\beta}^T\textbf{x}_i+\phi(\textbf{s}_i),
\end{equation}
where the intercept $\alpha$ and linear weights, $\boldsymbol{\beta}$ were given a zero mean Gaussian prior with variance 100. 
The spatial random effect $\phi(\textbf{s})$ followed the Barrier model of \citet{Bakka_etal:2019}, which is a non-stationary Gaussian process whose covariance does not travel through physical barriers -- islands in our application. 

The Barrier model is defined as a Stochastic Partial Differential Equation (SPDE) which leads to a Gaussian process with a non-stationary covariance function. For a stationary covariance function the covariance between two points $\s$ and $\s^\prime$ depends only on the distance $d(\s,\s^\prime)$. In the Barrier model the covariance travels only through water, while land areas are considered as barriers through which the covariance does not travel. This property is achieved by defining the spatial random effect, $\phi(s)$, as the solution to the following SPDE:
\begin{align}\label{eq:barrier_model}
 \phi(s) - \nabla \frac{r^2}{8} \phi(s) &= r\sqrt{\frac{\pi}{2}}\sigma_{\phi} \mathcal{W}(s), \text{ for } s \in \Omega_w \nonumber\\
 \phi(s) - \nabla \frac{r^2_b}{8} \phi(s) &= r_b \sqrt{\frac{\pi}{2}} \sigma_{\phi} \mathcal{W}(s), \text{ for } s \in \Omega_l, 
\end{align}
where $\Omega_w$ is water, $\Omega_l$ is land, $\mathcal{W}(s)$ is white noise, $\sigma_{\phi}$ is the marginal standard deviation of the process and $r$ is the range parameter governing how fast the correlation decreases with distance through water. The parameter $r_b$ is the correlation range on land and it is set to be a fraction of $r$ to remove the correlation there \citep{Bakka_etal:2019}. Here, we use the default value $r_b = 0.2r$. 
The details on implementing the barrier model are given in Appendix~B.1.
The parameters of the barrier model follow a Penalized Complexity (PC) prior \citep{Simpson2014a,Bakka_etal:2019}. These priors shrink the effect towards a base model, where $\sigma \rightarrow 0$ and $l \rightarrow \infty$. We defined the prior through probabilities $p(\sigma > 1) = 0.01$ and $p(l < 0.5km) = 0.01$.

As detailed in Section~\ref{sec:survey_data} the transect observations correspond to the number of pikeperch larvae caught with a net that samples a volume of water over a transect. 
We denote by $\s_i$ the coordinates of the middle point of the $i$th transect, by
$y_i$ the number of larvae caught over the transect and by $\x_i=\x(\s_i)$ the environmental covariates at the middle of the transect. 
Since survey transects were short compared to the resolution of the mesh used to implement the model (see Appendix~B.1), we treated the larvae density as a constant over each transect. 
Hence, the observation model for the survey data $\y=[y_1,\dots,y_n]^T$ is 
$p(\y|\boldsymbol{\lambda},\boldsymbol{\epsilon}) = \prod_{i=1}^n \mathrm{Poisson}(V_i\lambda_i\epsilon_i)$ where $V_i$ is the volume of water sampled, $\lambda_i=\lambda(\s_i,\x_i)$ is the intensity of the log Gaussian Cox process and $\boldsymbol{\epsilon}=[\epsilon_1,\dots,\epsilon_n]^T$ is a vector of independent random effects capturing overdispersion in larval counts arising from local phenomena and stochasticity in sampling.
 Note though that, in the survey data set, each point has the same sampling volume $V_i=V$ (28.35 m$^3$) so we gave Gamma prior for the random effects, $\epsilon_i \sim \mathrm{Gamma}(r,1/r)$, which leads to negative binomial observation model for the larval counts 
\begin{equation}\label{eq:survey_data_model}
 p(\y|\lambda,r) = \prod_{i=1}^n \mathrm{Negative-Binomial}(V_i\lambda_i,r)
\end{equation}
so that the observed larval counts are actually modeled to arise from an overdispersed log Gaussian Cox process which approaches the Poisson process as the overdispersion parameter $r$ increases \citep[see also,][]{Liu+Vanhatalo:2020}. We gave $\text{Gamma}(\sqrt{10},1/\sqrt{10})$ prior for the overdispersion parameter $r$.

\subsubsection{Model for larval presence}

As an alternative to larval abundance modeling we considered also an occurrence model where the occurrence of larvae in a survey sample was modeled with 
\begin{equation}\label{eq:survey_data_model_binom}
\mathbbm{1}(y_i>0)  \sim \text{Bernoulli}\left(\pi(\textbf{s}_i,\textbf{x}_i)\right)
\end{equation}
where $\mathbbm{1}(y_i>0)$ is an indicator function returning one if $y_i>0$ and zero otherwise and $\pi(\textbf{s}_i,\textbf{x}_i) = \text{logit}^{-1} \left(\alpha+\boldsymbol{\beta}^T\textbf{x}_i+\phi(\textbf{s}_i)\right)$ denotes the probability of presence for pikeperch larvae. The priors for model parameters $\alpha$, $\beta$ and $\phi$ were the same as in the abundance model.

\subsection{Joint model for expert assessments and survey data}\label{sec:joint_model}

\subsubsection{Model for expert opinions}

We constructed the model for expert assessments by first building a model for experts' subjective probabilities on the occurrence of larvae.
For this, we denote by $\bar{\pi}_{ji}\in[0,1]$ the $j$th expert's subjective probability that  larvae are present at location $s_i$ and model this probability with Beta distribution 
\begin{equation}
\bar{\pi}_{ji}\sim Beta(\bar{\mu}_{ji}, \bar{s}_j),
\end{equation}
where $\bar{\mu}_{ji}=E[\bar{\pi}_{ji}]$, and $\bar{s}_j$ is the (prior sample size) parameter governing the spread of the Beta distribution. We fixed $\bar{s}_j=2$ to encode vague prior predictive distribution for expert opinions. 
We then model the expected value of an expert's subjective probability with logistic regression so that it is a function of the log of the true larval density or the logit of the true probability of presence of larvae, so that
\begin{equation}\label{eq:expert_assessment_jointModel_part2}
\bar{\mu}_{ji} = \mathrm{logit}^{-1}(\bar{\alpha}_j + \bar{c}_j(\boldsymbol{\beta}^T\textbf{x}_i+\phi(\textbf{s}_i)) + \bar{\varphi}_{j}(\textbf{s}_i)). 
\end{equation}
Here, the parameter $\bar{\alpha}_j$ is an intercept, $\bar{c}_j$ is a parameter for the expert's skill and $\bar{\varphi}_{j}(\textbf{s}_i)$ is a residual error. Note that, for example,  $\log\lambda_i - \alpha=\boldsymbol{\beta}^T\textbf{x}_i+\phi(\textbf{s}_i)$, so $\text{logit} (\mu_{ji})$ is proportional to $\log\lambda_i$ (or $\mathrm{logit}(\pi_i)$) but we have subtracted $\alpha$ from \eqref{eq:expert_assessment_jointModel_part2} in order to improve the identifiability of the parameters. The parameter $\bar{c}_j$, thus, describes how strongly an expert's assessment follows the true underlying larvae intensity or probability of presence, so that a positive $\bar{c}_j$ indicates that an expert has information about larvae distribution. 
The error terms $\bar{\varphi}_{j}(\textbf{s}_i)$ correct for spatially correlated local biases in the expert assessment that can result from actual bias in an expert's opinion but also from inaccuracies in expressing them. For example, it is likely that the experts coloured the maps with smaller resolution than the true resolution in the larvae presence/absence pattern, so $\bar{\varphi}_{j}(\textbf{s}_i)$ also accounts for spatial correlation resulting from this.

In case of a conflict between expert assessment and survey data, we want the model to explain the expert assessment with $\bar{\varphi}$ instead of $\bar{c}_j(\boldsymbol{\beta}^T\textbf{x}_i+\phi(\textbf{s}_i))$ so that an expert's assessment does not provide information on larvae intensity or occurrence probability.
To attain this behaviour we give relatively stricter prior for $\bar{c}_j$ than to $\bar{\varphi}$.
Hence, priors for the parameters were  $\bar{\alpha}_{j}\sim N(0,2^2)$ and 
$\bar{c}_j \sim N(0,0.5^2)$. The latter implies 95\% prior probability that $\bar{c}_j$ is less than one. 
We then modeled the expert bias function $\bar{\varphi}_{j}(\textbf{s}_i)$, with Besag-York-Mollié (BYM) -model \citep{besag1991bayesian}, which induces spatial correlation among neighboring pixels in an expert assessment graph (see Appendix~B.1 for details).
In the BYM-model $\bar{\varphi}_{j}\sim N(0,Q^{-1})$, where the precision matrix $Q = \tau_u R +\tau_v I$, $I$ is the identity matrix corresponding to the i.i.d. random effect, and
\begin{equation}
 R_{kl} = \begin{cases} n_k, &k=l\\
 						 -\mathbbm{1}\{k\sim l\}, &k\neq l	\end{cases}
\end{equation}
where $n_k$ is the number of neighbors of vertex $k$, and $k \sim l$ indicates that vertices $k$ and $l$ are neighbors, i.e. they are connected by an edge.
The hyperparameters of the BYM model are precision for the spatially correlated effect $\tau_u$ and precision for the i.i.d. effect $\tau_v$. We gave both of them gamma prior $\Gamma(2,8)$. When marginalizing over the precision this induces a heavy tailed scaled Student-$t$ 
distribution with $\nu=4$ and $s=2$ for $\bar{\varphi}$.

\subsubsection{Likelihood functions for the expert assessments}

As explained in Section~\ref{sec:expert_elicitation} the expert assessments were coded as raster maps so that the assessment of the $j$'th expert for grid cell $i$ at location $\s_{i}$ is a categorical variable $z_{ij}\in\{1,2,3,4\}$ if the grid cell belongs to the expert's assessment region (Figure~A.2). If the grid cell is outside expert's assessment area $z_{ij}=\text{NA}$, which means that the corresponding grid cell is excluded from the likelihood function of the expert's assessments to be detailed below. We ordered the categories so that $z_{ij}=1$ corresponds to the lowest ($<10\%$) and $z_{ij}=4$ corresponds to the highest ($>90\%$) subjective probability of presence of an expert. 

As the first model for the expert assessments we treated them as binary statements about presence of pikeperch larvae. In this model, a location within an expert's assessment region was considered to be labeled as presence if the expert had given it over 50\% probability for larvae presence (i.e. $z_{ij}\in \{3,4\})$) and otherwise it was considered as an absence. 
The observation model for the binary absence assessment is then
\begin{align}\label{eq:binary_expert_assessments}
\text{Pr}(z_{ij}\in \{1,2\})&=\text{Pr}(\bar{\pi}_{ji}\leq 0.5)=F_{\text{Beta}}(0.5|\bar{\mu}_{ji},\bar{s}_j) 
\end{align}
where $F_{\text{Beta}}(0.5|\mu_{ji},s_j)$ is the cumulative distribution function of the Beta distribution. Similarly for the binary presence assessment we have $\text{Pr}(z_{ij}\in \{3,4\})=1-F_{\text{Beta}}(0.5|\bar{\mu}_{ji},\bar{s}_j)$ 
This observation model induces a likelihood function for $\text{logit}(\bar{\mu}_{ji})$ that resembles closely the logit-function. 
We can similarly form the observation model for all expert assessment categories
\begin{align}\label{eq:multiple_categories_likelihood}
\text{Pr}(z_{ij}|\mu_{ji},s_j)&= \begin{cases}
F_{\text{Beta}}(0.1|\mu_{ji},s_j), &\text{ if } z_{ij} =1 \\
F_{\text{Beta}}(0.5|\mu_{ji},s_j)-F_{\text{Beta}}(0.1|\mu_{ji},s_j), &\text{ if } z_{ij} =2 \\
F_{\text{Beta}}(0.9|\mu_{ji},s_j)-F_{\text{Beta}}(0.5|\mu_{ji},s_j), &\text{ if } z_{ij} =3\\
1-F_{\text{Beta}}(0.9|\mu_{ji},s_j), &\text{ if } z_{ij} =4,
\end{cases}
\end{align}
which again induces likelihood functions for $\text{logit}(\mu_{ji})$ and through that to the model parameters. 

We made a further assumption that, conditionally on the model parameters, the expert assessments are mutually independent. Hence, the joint distribution for all expert assessments over the whole study region can be written as
\begin{equation}
p\left(\z_1,\dots,\z_{J}|\bar{\alpha},\bar{c},\boldsymbol{\beta},\varphi,\bar{\varphi}\right) = \prod_{j=1}^{J} \prod_{i=1}^{n_j} p(z_{ij}|\bar{\alpha}_j,\bar{c}_j,\boldsymbol{\beta},\varphi,\bar{\varphi}_j)
\end{equation}\label{eq:exp_beta_cdf}
where $J$ is the number of experts, $n_j$ is the number of grid cells inside the assessment area of the $j$th expert, $\bar{\alpha}=\{\bar{\alpha}_1,\dots,\bar{\alpha}_J\}$, $\bar{c}=\{\bar{c}_1,\dots,\bar{c}_J\}$ and $\bar{\varphi}=\{\bar{\varphi}_1,\dots,\bar{\varphi}_J\}$.
When we combine the two data sets, $\y$ (the survey observations) and $\z_j, j=1,\dots,J$ (the expert observations), the observations are independent given the latent function and model parameters so that, in case of count observations from surveys, the full observation model is 
\begin{equation}
p(\y,\z_1,\dots,\z_{J}|\bar{\alpha}, \bar{c},\bar{\varphi},\alpha,\boldsymbol{\beta},\varphi,r) = \left[\prod_{i=1}^n p(y_i|\lambda_i,r)\right]\prod_{j=1}^{J}\prod_{k=1}^{n_j} p(z_{kj}|\bar{\alpha}_j, \bar{c}_j,\bar{\varphi}_j,\boldsymbol{\beta},\varphi).
\end{equation}
The full model with the presence-absence model for the survey observations is constructed analogously.

\section{Posterior inference and model comparison}

We implemented all the models and conducted the posterior inference using the Integrated Nested Laplace Approximation (INLA) R package \citep[R-INLA,][]{rue2009approximate}. 
The technical implementation of the barrier model requires a triangular mesh over the study area on which the model is constructed \citep{Bakka_etal:2019}. 
The survey observations and expert assessments are then mapped to the mesh with projection matrices. 
The details on how we constructed these are given in the Appendix~B.1.
R-INLA does not support the exact likelihood functions derived in Section~\ref{sec:joint_model}, so we searched for the closest match to them from among the Binomial likelihood functions, which are readily available in R-INLA. The details of the matching process and the resulting likelihood functions are given in the Appendix~C. 
After setting up the models, we constructed the INLA approximation for the model parameters and used the model to predict the larvae intensity over the whole study area.

Since the survey data were collected with a standardized and extensively tested sampling method \citep[see][and their references]{Veneranta_etal:2011,Kallasvuo_etal:2017} and the sampling time was matched with the peak larval time \citep{Liu+Vanhatalo:2020}, we considered them to accurately reflect the larval areas of pikeperch. On the other hand, we had less evidence on the expert elicitation process and the reliability of the experts. Hence, we considered the survey data to be more reliable source of information on larvae distribution than the expert assessments, and compared the alternative models' based on how well they predict survey data that have been left out from the training data.
The INLA approximation gives access to approximate leave-one-out (LOO) predictive densities for all observations. These are also called Conditional Predictive Ordinate (CPO) values and, for the $i$'th survey observation, it is defined as
\begin{equation}
\mathrm{CPO}_i = p(y_i|\mathbf{x}_i,\mathbf{s}_i,D_{\setminus i})
\end{equation}
where $D_{\setminus i}$ includes all the other data except $\{y_i,\mathbf{x}_i,\mathbf{s}_i\}$. The CPO values can be used to calculate the widely used average LOO cross-validation log predictive density
\begin{equation}
\mathrm{lpd} = \frac{1}{n_{\mathrm{survey}}}\sum_{i=1}^{n_{\mathrm{survey}}} \log \mathrm{CPO}_i.
\end{equation}

With models where survey data are modeled as presence-absence data, we can calculate three additional model comparison metrics. First, we calculated classification accuracy, 
$\mathrm{ACC} = \frac{1}{n_{\mathrm{survey}}}\sum_{i=1}^{n_{\mathrm{survey}}} \mathbbm{1}(\mathrm{CPO}_i\geq 0.5)$, i.e. the ratio of correctly classified observations.
Because the observations are skewed towards absences, we also calculated the balanced classification accuracy $\mathrm{bACC} = \frac{\mathrm{TPR}+\mathrm{TNR}}{2}$, which is the average of true positive rate (TPR) and true negative rate (TNR) that are calculated similarly to ACC.
Log predictive density is known to be sensitive to outliers (i.e., observations for which the predictive density is much lower than in average) for which reason \citet{CRPS_2007} proposed to use the Continuous Ranked Probability Score (CRPS) as an alternative metric for comparing probabilistic predictions. The CRPS for a single prediction is defined as
\begin{equation}\label{eq:crps}
\text{CRPS}(F_i,y_i) = E_{F_i}|Y_i-y_i| -\frac{1}{2}E_{F_i}|Y_i-Y_i^\prime|,
\end{equation}
where $F_i$ is the posterior predictive distribution given by the model being evaluated, $Y_i$ and $Y_i^\prime$ are independent and identical random variables that follow $F_i$ and $y_i$ is an observation. In the case of Bernoulli likelihood for survey data, the LOO posterior predictive distribution for the $i$'th observation is  $Y_i \sim F = \text{Bernoulli}(\hat{\pi}_i)$ where $\hat{\pi}_i=\text{Pr}(Y_i=1|\mathbf{x}_i,\mathbf{s}_i,D_{\setminus i})$.  Hence, we can simplify the CRPS for the LOO posterior predictive distributions of these models to the square of the posterior predictive probability of the incorrect class:
\begin{align*}
\text{CRPS}(F_i,y_i) &= \hat{\pi}_i(1-y_i)+\left(1-\hat{\pi}_i\right)y_i -\frac{1}{2}\left(2\hat{\pi}_i\left(1-\hat{\pi}_i\right)\right) \\
 			&= \begin{cases}
 			  \hat{\pi}_i^2 & \text{ if } y_i=0 \\
 			  (1-\hat{\pi}_i)^2 & \text{ if } y_i= 1
 			  \end{cases}\\
 			  & = (1-\mathrm{CPO}_i)^2.
\end{align*}


\section{Results}

We received in total 11 expert assessments out of which ten were used for the analyses. 
The assessed areal coverage of experts varied considerably. Some experts expressed their views for very small regions whereas some of them covered almost the entire study area (see Figure~A.2). 
One expert assessment was excluded, as the area assessed by the expert was considered too small for the analysis. 
In some cases, the request of first drawing the expert’s own assessment area in the map was poorly understood and needed to be clarified later on.
Some experts were more uncertain in their statements (they used only categories 2 or 3) than others (who used all four categories). 
In general the digitalization of the expert assessment maps worked reasonably well but in few cases borders between two colors were hard to distinguish and needed some interpretation by us.

In general, the models integrating survey data and expert assessments (hereafter survey+expert models) outperformed the survey only model in their predictive performance (table~\ref{table:model_comparison}). In occurrence predictions, that is survey data were modeled as occurrence data (Bernoulli likelihood \eqref{eq:survey_data_model_binom}), the best model was the one where expert assessments were modeled as presence-absence statements \eqref{eq:binary_expert_assessments}.
However, the survey+expert models outperformed the survey only model in terms of classification accuracy (ACC and bACC) and CRPS statistics but not in LPD statistics (see table~\ref{table:model_comparison}).
Since LPD is more sensitive to outlying observations than CRPS \citep{CRPS_2007}, this indicates that survey+expert models made more over-confident false presence-absence predictions than the survey only models.
In count predictions, that is survey data were modeled as count data (Negative-Binomial likelihood~\eqref{eq:survey_data_model}), the survey+expert models had clearly better LPD statistics than the survey only model (table~\ref{table:model_comparison}). The differences in LPD statistics between alternative larval abundance models were rather large and the best model was the survey+expert model where we used all four expert assessment categories, observation model \eqref{eq:multiple_categories_likelihood}.

\begin{table}[t]\caption{Posterior predictive model comparison for alternative models. First two columns specify the observation model for survey data and expert elicited maps; "--" in the latter denotes a survey data only model, "p/a" denotes the presence absence model (\eqref{eq:survey_data_model_binom} for survey data and \eqref{eq:binary_expert_assessments} for expert assessments), and "abu" denotes abundance model \eqref{eq:survey_data_model} for survey data and four categories model \eqref{eq:multiple_categories_likelihood} for expert assessments. Other abbreviations are lpd for leave-one out cross validation log predictive density, ACC for classification accuracy, bACC for balanced classification accuracy and CRPS for conditional predictive ordinate.}\label{table:model_comparison}
\centering
\begin{tabular}{c c c c c c}
\hline
\multicolumn{2}{c}{Observation models} & lpd & ACC & bACC & CRPS \\
Survey data & Expert maps & & & & \\
\hline
p/a & -- & -83.6 & 0.718 & 0.683 & 0.538\\
p/a & p/a & -95.2 & 0.737 & 0.724 & 0.436\\
p/a & abu & -87.3 & 0.737 & 0.721 & 0.465 \\
\hline
abu & -- &  -452 & - & - & -\\
abu & p/a & -410 & - & - & -\\
abu & abu & -294 & & & \\
\hline
\end{tabular}
\end{table}

\begin{figure}[t]
\begin{subfigure}[b]{\textwidth}
	\hspace{0cm}\includegraphics[scale=0.65]{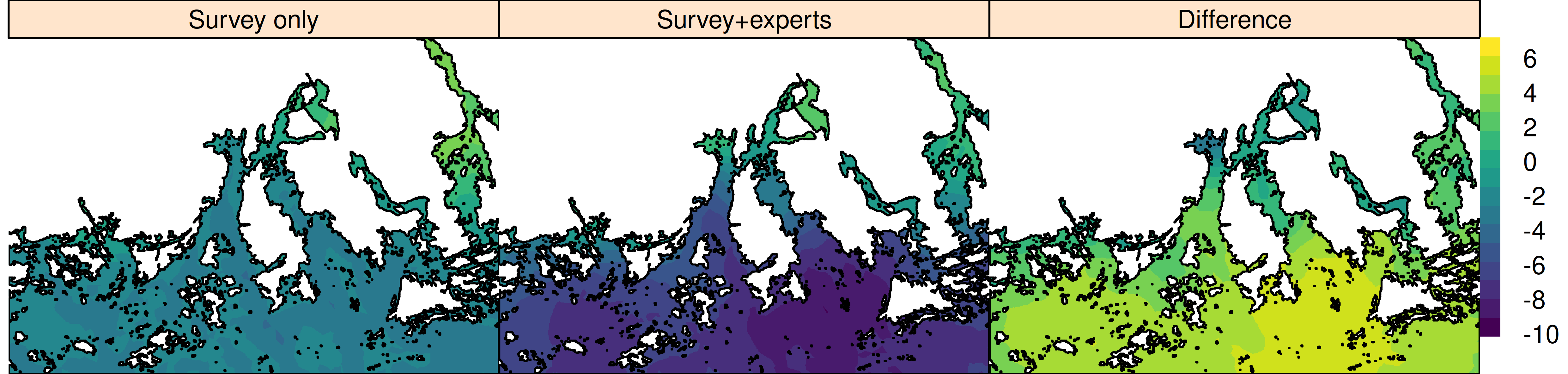}
\end{subfigure}
\begin{subfigure}[b]{\textwidth}
	\hspace{0cm}\includegraphics[scale=0.65]{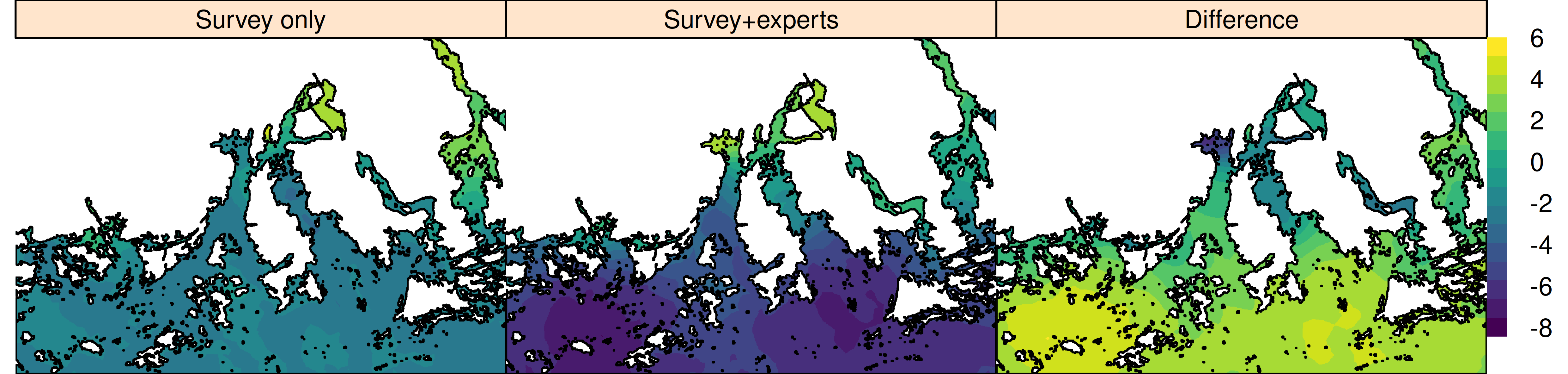}
\end{subfigure}
\caption{Posterior predictions for larval distribution in the study area with survey only and survey+experts models and their difference ([survey only] - [survey+experts]). On top row, the survey data are modeled as occurrence data and the maps show the posterior predictive mean of the log odds ratio for larval occurrence and their difference. On bottom row, survey data are modeled as count data and the maps show the posterior predictive mean of the log larval density and their difference. In both rows, the expert assessments included all four categories.}\label{fig:predictive_density_betaExpert}
\end{figure}

\begin{figure}[t]
\begin{subfigure}[b]{\textwidth}
	\hspace{0cm}\includegraphics[scale=0.65]{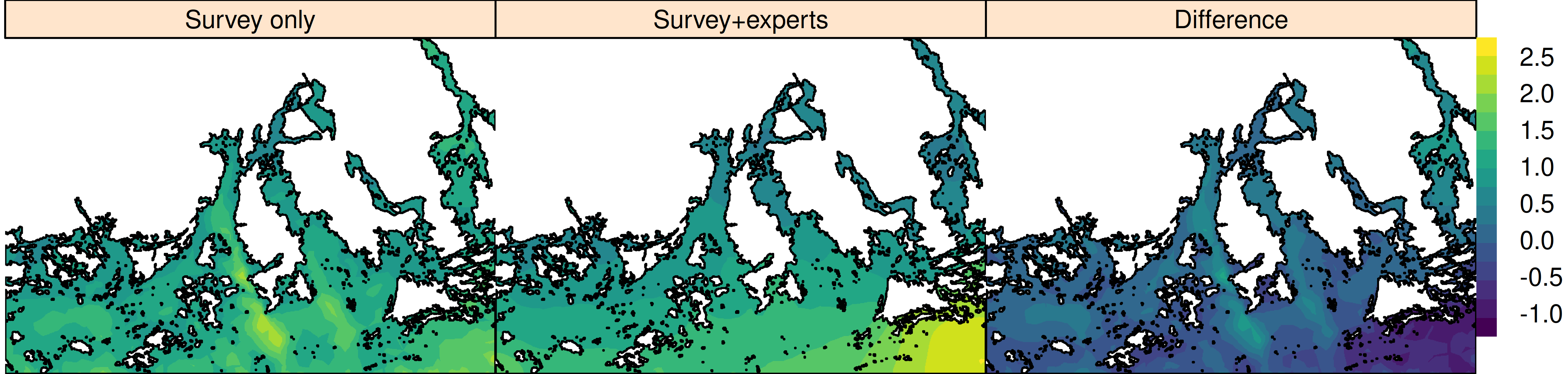}
\end{subfigure}
\begin{subfigure}[b]{\textwidth}
	\hspace{0cm}\includegraphics[scale=0.65]{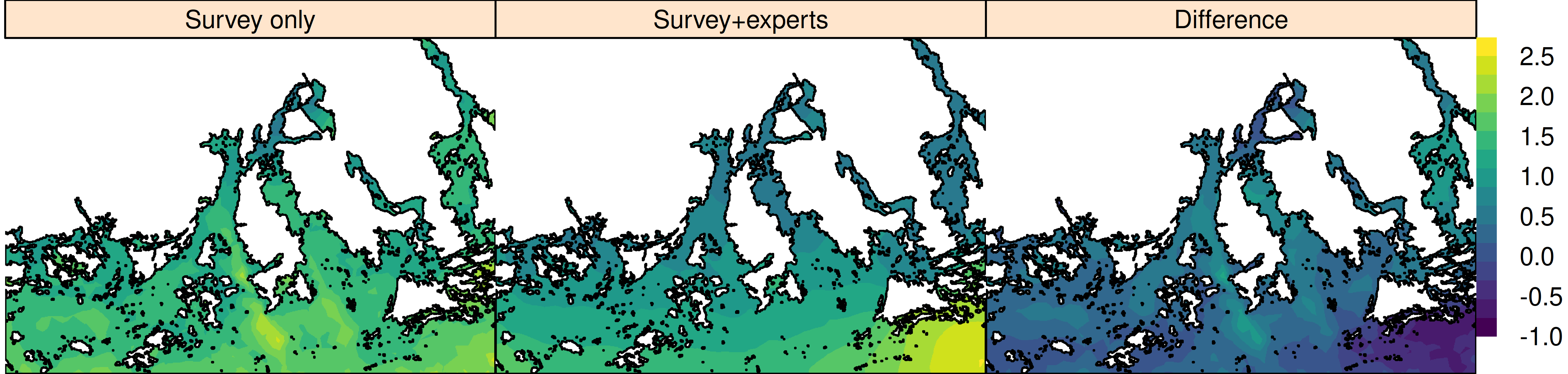}
\end{subfigure}
\caption{Posterior predictive uncertainty for larval distribution in the study area with survey only and survey+experts models and their difference ([survey only] - [survey+experts]). On top row, the survey data are modeled as occurrence data and the maps show the posterior predictive standard deviation of the log odds ratio for larval occurrence and their difference. On bottom row, survey data are modeled as count data and the maps show the posterior predictive standard deviation of the log larval density and their difference. In both rows, the expert assessments included all four categories. }\label{fig:predictive_density_betaExpert_sd}
\end{figure}

All the models produced posterior predictive larval distribution maps that were similar in their overall pattern; that is, the larvae density and occurrence probability is the highest in the sheltered bays in the northern parts of the study area and smallest in open sea areas in the south (see figures~\ref{fig:predictive_density_betaExpert} and~B.4a). However, there are clear differences in the finer scale patterns of the predictions. For example, the survey+expert models predict consistently lower densities and occurrence probabilities in southern areas than the survey only models. 
The survey+expert models predicted also somewhat higher larval densities and occurrence probability than the survey only model in some of the bays in the north whereas in the north-easternmost bay the survey only models tended to predict higher densities and occurrence probabilities than the survey+expert model. 
In general, the survey+expert models reduce the uncertainty compared to survey only models almost everywhere in the study area. Only in the south-eastern and south-western areas, the survey+expert models had larger posterior predictive standard deviation for the log intensity and logit probability than the survey only models (see figures~\ref{fig:predictive_density_betaExpert_sd} and B.4b).

\begin{figure}
\begin{subfigure}{0.45\textwidth}
  \includegraphics[width=\linewidth]{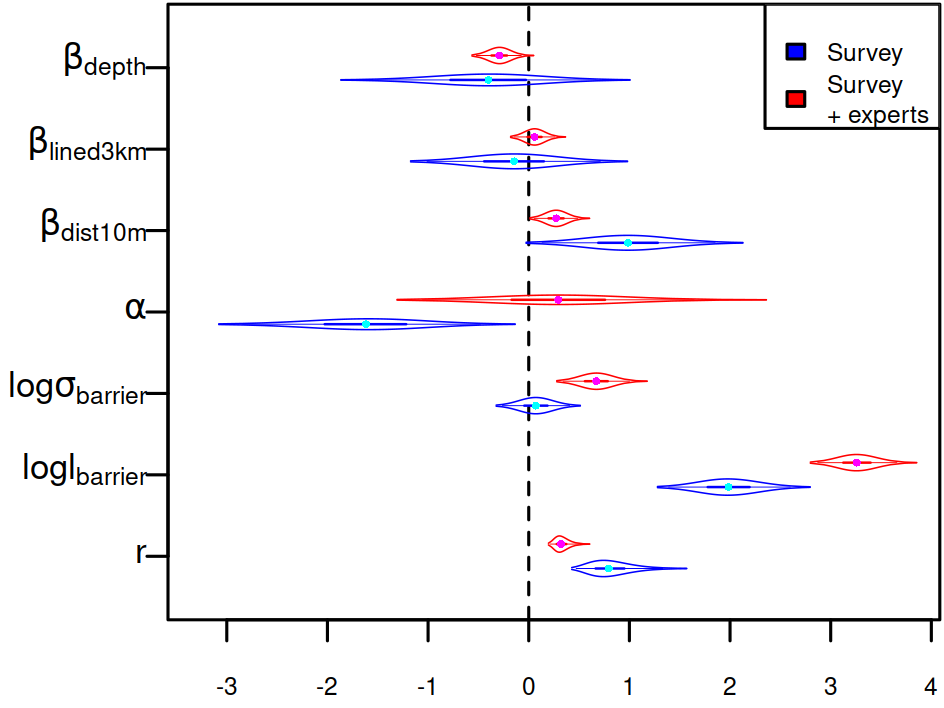}
  \caption{Survey model parameters}\label{fig:nbin_beta_violin}
\end{subfigure}
\begin{subfigure}{0.45\textwidth}
		\includegraphics[width=\linewidth]{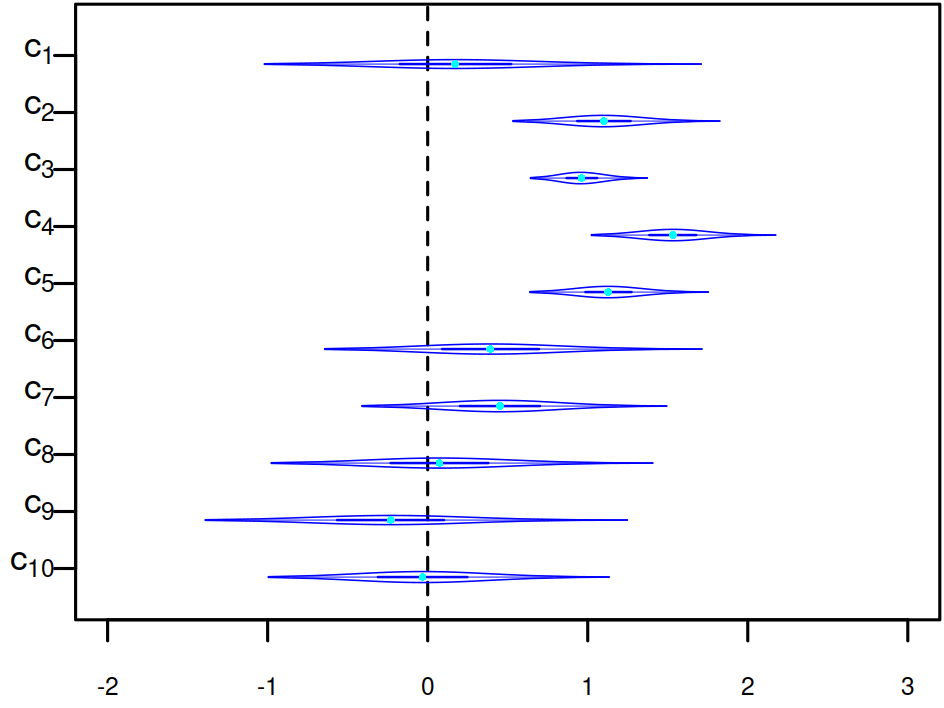}
		\caption{Expert coefficients}\label{fig:nbin_beta_coeff}
\end{subfigure}
\caption{Posterior distributions for model parameters in the survey+experts model with Negative-Binomial observation model for survey data and Beta observation model for expert assessments.}\label{fig:parameter posteriors}
\end{figure}

Experts' reliability, when measured by the $\bar{c}$ parameters, varied considerably (see figures~\ref{fig:nbin_beta_coeff} and~B.2). Experts 2-5 were the reliable ones having significantly positive $\bar{c}$ parameters whereas all the other experts were unreliable in the sense that the posterior distributions their $\bar{c}$ parameters did not differ from zero significantly. 
The third expert showed some local bias in his/her assessment in a couple of inner bayes in the north but otherwise the spatial bias terms of the reliable experts were small (see Figure~B.3). 
The posterior distributions of the parameters of the survey data model were considerably narrower in the survey+expert models than in the survey only models (figures~\ref{fig:nbin_beta_violin} and~B.1). 
Log larval density and log odds ratio of larval occurrence responded negatively to depth and openness (lined3km) and positively to distance to 10m depth curve. The overdispersion in survey data was significant since the posterior distribution of overdispersion parameter, $r$, was concentrated in small values. The barrier model had more significant effect on log larval density and log odds-ratio of occurrence in survey+expert models than in the survey only models 

We also conducted sensitivity analyses for the prior distributions.
With too relaxed barrier model priors, INLA sometimes ran into numerical problems, with the latent variable $\eta$ evaluating to \verb!NAN!- or \verb!INF!-values at several points in the mesh. 
Numerical problems did not arise when we defined the PC-priors to correspond the general knowledge on likely scale and spatial correlation distance in log density and log odds ratio (the priors reported in Section~\ref{sec:survey_data}). 
When the BYM-model was left out, or when we restricted its variance parameters too heavily towards zero by prior distributions, the survey+expert models effectively ignored the survey data information and fitted the log abundance and log odds ratio based on experts information only.
However, when expert bias terms (BYM-model components) were \emph{a priori} allowed to vary more than the log density or log odds ratio, conflict between survey data and expert assessment was explained by the bias term, and the expert assessment did not contribute to the posterior distribution of log density and log odds ratio in practice. 
In addition to prior distributions, the resolution of expert-mesh had an impact on how much weight we gave to expert assessments so that larger grid cells in the expert-mesh implied reduced the effect of expert assessments compared to fine grid cells.


\section{Discussion}

Our results show clear differences between the survey only and survey+expert models both in predictions and parameter inference. 
In general, the survey+expert models outperformed the survey only models in their predictive performance, indicating that adding expert information to SDMs can be benefitial. 
The general pattern of the predicted species distributions was similar across different models, but there were clear differences in some areas where expert assessments were available (Figure~\ref{fig:predictive_density_betaExpert}). 
The biggest differences between the alternative models being in the south.
This is reasonable since all experts, who had marked the southern areas into their assessment regions, had given the smallest probability for larval occurrence there (see Figure~A.2) whereas there were only a few survey observations from the southern part of the study region. 
Hence, the expert assessments have the largest impact in the south. 
In addition to changing the predictive mean of the larvae abundance and occurrence probability, including expert assessments into the models decreased their predictive uncertainty as well.
Similarly, the posterior distributions of the parameters of the larval density and occurrence probability models were considerably narrower in the survey+expert models than in the survey only models (figures~\ref{fig:nbin_beta_violin} and~B.1). 
All these results show that the expert assessments provided significant information on factors affecting larvae distribution.

These results suggest that expert elicitation can be a viable solution in studies concerning distribution of species in space and time when resources for scientific sampling are limited. However, our results demonstrate also that expert elicitation should be applied with care and optimally combined with more objective information to reduce the effect of subjective biases. These findings align well with earlier studies that have shown that the skill of experts in assessing species distributions varies and depends, for example, on study area, species and background of an expert \citep{Pearce+etal:2001,DiFebbraro+etal:2018,Crawford+etal:2020}. Bias correction and weighing expert assessments based on experts' skills have been emphasized also in many other expert assessment tasks \citep{OHagan+etal:2006,Burgman+etal:2011,Dias_etal:2018,perala2020}.

The expert elicitation process turned out well-functioning. Important first step was to confirm the commitment of the experts to the process.  As a result, responses were gotten from all candidate experts. 
For successful implementation of an expert elicitation by post, it is important that instructions for filling questionnaire are clear and that the level of details are kept meaningful. Another, but more consuming option, would be to run a face-to-face meeting or an expert elicitation workshop to collect the expert’s views.

The technical implementation of our models was done using the R-INLA software \citep{rue2009approximate} since it allows straightforward implementation of the barrier model~\eqref{eq:barrier_model}. The barrier model is justified in our application area since it is scattered by islands, peninsulas and other physical obstacles for marine species that make traditional stationary spatial random effect an unrealistic assumption \citep{Bakka_etal:2019}. 
This choice did not come without price, though, since we were restricted to the built-in likelihood functions in INLA which affected our practical choice for the expert assessment models (see Section~\ref{sec:joint_model} and Appendix~C). 
We believe this choice had only minor effect on the results though. 
More fundamental technical consideration is related to the prior distributions. Firstly, the priors for the hyperparameters of barrier model had to be chosen carefully to keep the R-INLA calculations numerically stable. 
Secondly, to prevent expert assessments from overruling the survey observations, the mesh resolution and the prior distributions for the hyperparameters of the expert bias terms (BYM-model parameters) had to be chosen reasonably. 
Too small mesh size and too narrow prior for the BYM-model prevents the model from rejecting biased expert assessments. 
This is understandable because we included the expert assessments into our models as point-wise observations at expert-mesh nodes.
 Hence, if expert bias terms (BYM-model components) were restricted to near zero, the combined likelihood of expert assessments would outweight the likelihood of survey data in information concerning the log larvae density or log odds ratio of larvae occurrence probability (there are far more expert-mesh nodes than survey data observation locations).
For this reason, the expert-mesh resolution should be chosen so that it corresponds to the resolution at which the experts can draw their maps.
Similarly, the variance parameters of the BYM-model should be given wide priors.

In summary, our results are encouraging for further development of expert elicitation methods in species distribution modeling. They suggest also that the methods proposed in this work could also be scaled to larger scale applications; such as environmental accounting and environmenal management. 
Biodiversity loss has become an equally important challenge for humanity and wellbeing of our planet as climate change. 
Today, confrontation between nature and economic wellbeing has started to disappear and general demand for introducing natural capital into national accounting systems to appear \citep{Dasgupta:2021}. This demand parallels other calls for biodiversity preserving measures \citep[e.g., the System of Environmental Economic Accounting][]{SEEA_EA:2021} in that they all rely on monitoring of species, biodiversity and ecosystem processes.
Current technology does not, however, allow continuous measuring of biota over space and time so predictive models are needed to fill in the gaps. 
Species distribution models are routinely used for this task so that they are trained with observational data to make predictions on species occurrence and abundance at spatiotemporal locations not covered by observations. However, survey data can be expensive and logistically challenging to collect whereas expert information can often be collected with relatively inexpensively. For this reason expert elicitation is a tempting method to collect compelementary information to survey data \citep{pearman2020predicting,murray2009useful}. The other benefit from expert elicitation in the context of environmental accounting could be stakeholder engagement since management and policy decisions are typically better received by stakeholders and other interest groups if they have been heard and their knowledge incorporated in the decision making process \citep{LaMere_etal:2020b}.

\section{Conclusions}

We have presented a formal Bayesian approach to include expert information into species distribution modeling. 
Our approach effectively integrates spatial expert information with point-wise survey data and allows expert calibration and assessment of experts' reliability.
Our results show that expert information can significantly improve species distribution predictions compared to predictions conditioned on survey data only. 
This suggests that expert elicitation can be an efficient tool, for example, in natural resources management, conservation area planning and in other applications that require knowledge on distributions of species and other biotic variables. 
However, our results also demonstrate that experts' reliability may considerably vary, and that even generally reliable experts can have spatially structured biases in their beliefs.
Hence, expert information should be analyzed with care and preferably calibrated with more objective information sources such as survey data.


\section*{Acknowledgements}

This work has been funded by the Academy of Finland (grant 317255, KK and JV) and the Finnish Operational Program for the European Maritime and Fisheries Fund (EMFF, AL and SK).

\bibliographystyle{ba}
\bibliography{references}









\end{document}



\begin{frontmatter}

\title{Supplementary Material: Species Distribution Modeling with Expert Elicitation and Bayesian Calibration}
\runtitle{SDMs with Expert Elicitation and Bayesian Calibration}

\begin{aug}
\author{\fnms{Karel} \snm{Kaurila}\thanksref{addr1}},
\author{\fnms{Sanna} \snm{Kuningas}\thanksref{addr2}},
\author{\fnms{Antti} \snm{Lappalainen}\thanksref{addr2}},
\and
\author{\fnms{Jarno} \snm{Vanhatalo}\thanksref{addr1,addr3}}

\runauthor{}

\address[addr1]{Department of Mathematics and Statistics, University of Helsinki}
\address[addr2]{Natural Resources Institute Finland}
\address[addr3]{Organismal and Evolutionary Biology Research Programme, University of Helsinki}


\end{aug}

\end{frontmatter}

\appendix
\counterwithin{figure}{section}

\section{Expert assessment maps}
\begin{figure}[hb]
		 \vspace{-0.5cm}
         \centering         
         \includegraphics[width=0.67\textwidth]{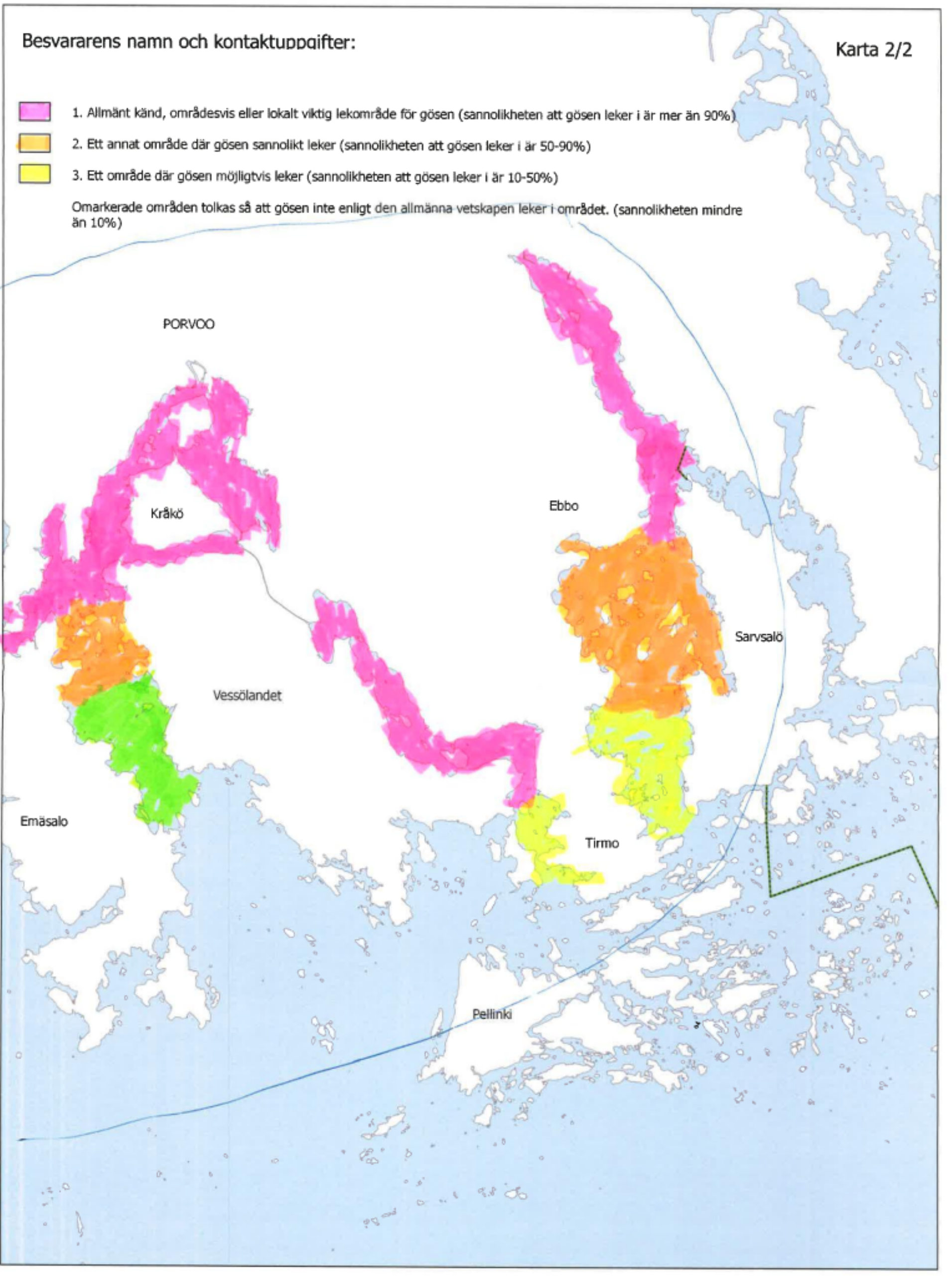}
\caption{An expert assessment map. Inside the hand drawn blue lines is the area where expert was confident to give their assessment. Sea areas outside this area were considered as missing information from this expert. This expert has coloured with pink the areas where they considered spawning to happen with over 90\% probability, with orange 50--90\% probability and with yellow or green 10--50\% probability. The uncolored areas inside expert's assessment region correspond to less than 10\% probability.}\label{fig:expertAssessment_map}
\end{figure}

\begin{sidewaysfigure}
    \centering
    \includegraphics[scale=0.7]{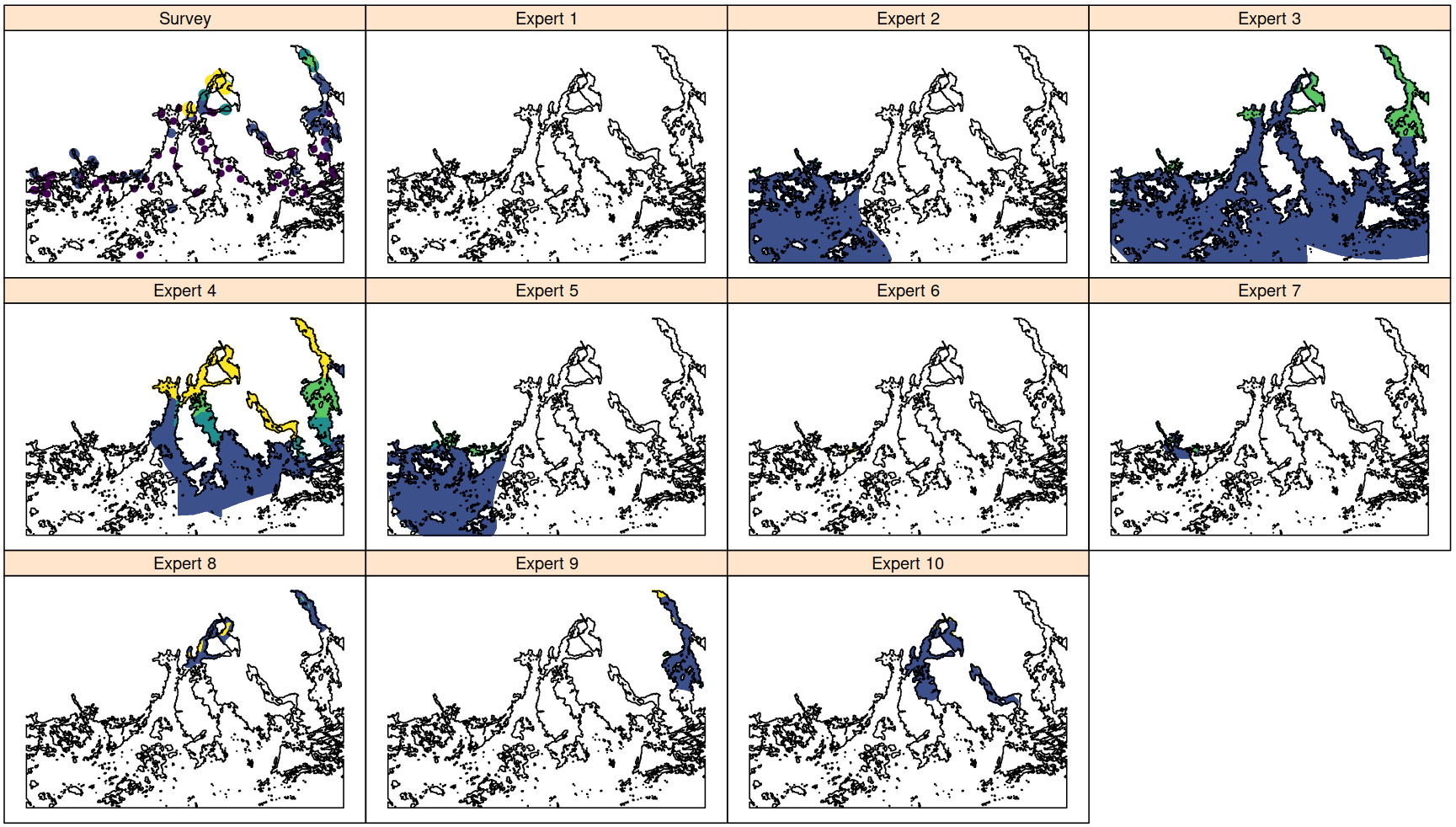}
    \caption{The visualization of the survey data (top-left corner) and the expert assessment maps after the digitalization (rest of the plots). The survey locations are drawn with dots whose colour represent the number of larvae caught at that site so that the deepest blue represents zero count and the log of the positive counts has the same color scale as in the lower row of Figure~3. In the expert assessment maps white denotes NA, dark blue denotes category 1, dark green category 2, light green category 3, and yellow category 4.}\label{fig:expert_observations}
\end{sidewaysfigure}

Figure~\ref{fig:expertAssessment_map} shows an example on expert drawn map. Expert opinions were digitized as rasters maps with the cell size matching the size of the environmental covariate maps, that is $50m \times 50m$, shown in Figure~\ref{fig:expert_observations}.

\section{Details on implementing the models in R-INLA}

\subsection{Building the model over triangular meshes}\label{sec:Barrier_mesh}

The coordinates of the data were in the EUREF-FIN coordinate system, which is a Finnish realization of the European Terrestial Reference System 1989 (ETRS89). EUREF-FIN uses the Universal Transverse Mercator (UTM) projection system, where Finland lies in the zone 35, and the Geodetic Reference System 1980 (GRS80). 

\begin{figure}[th]
    \begin{subfigure}[b]{0.49\textwidth}
         \includegraphics[scale=0.86]{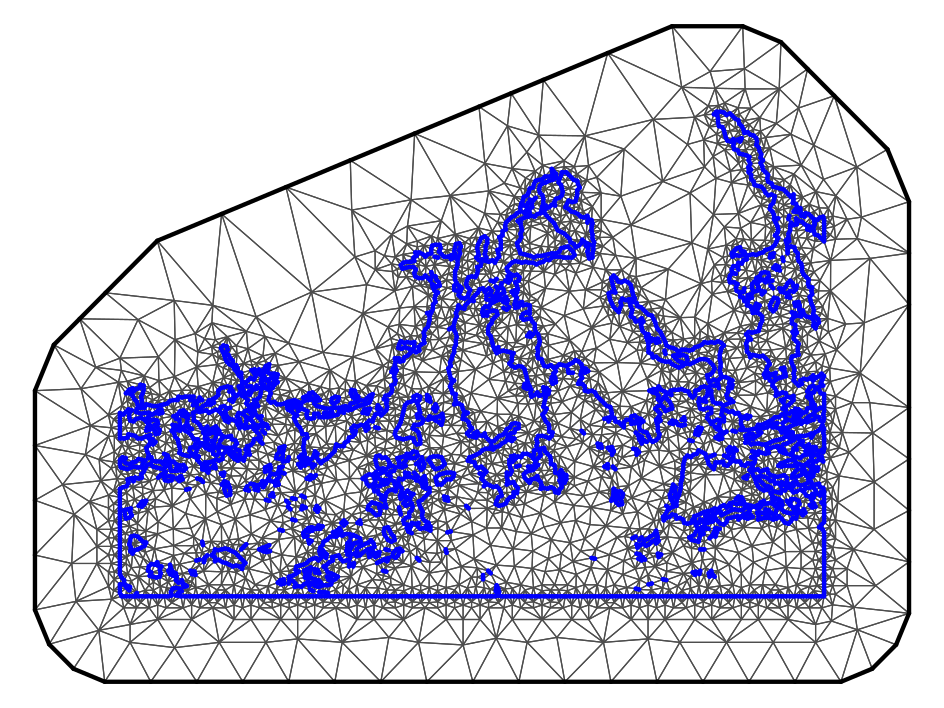}
         \caption{Barrier model mesh}
         \label{fig:model_mesh}
     \end{subfigure}
         \begin{subfigure}[b]{0.49\textwidth}
		 \includegraphics[scale=0.86]{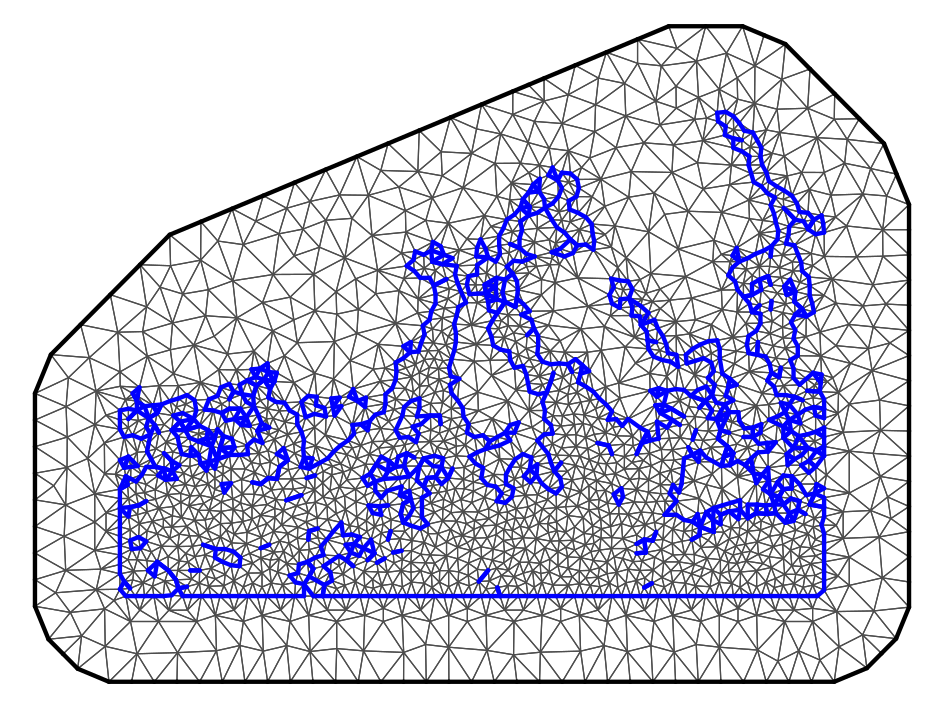}
         \caption{Expert model mesh}
         \label{fig:expert_mesh}
     \end{subfigure}
\caption{The triangulation meshes used in the barrier model (a) that was used as a spatial random effect in the larvae density model, and in the BYM model (b) that was used in the expert bias modeling.}
\end{figure}

The technical implementation of the barrier model requires a triangular mesh over the study area onto which the survey observations and expert assessments are then mapped with projection matrices \citep{rue2009approximate,martins2013bayesian,lindgren2011explicit,Bakka_etal:2019}. 
The triangular mesh for the barrier model was constructed over the whole study area so that the mesh captures the shape of the shorelines in the area in reasonable enough accuracy and so that the nearest survey observations would still be spatially separated in the mesh approximation. The mesh was constructed by using the R-INLA package. 
First we constructed a boundary polygon, which was constructed from the environmental covariate rasters with the \verb!rasterToPolygons! function in the \verb!raster! R package. 
The mesh was constructed by triangulating the study area using Delaunay Triangulation, where the study area is covered in triangles, whose angles are as close to $60^{\circ}$ as possible \cite{oloufa1991triangulation}. 
We used the following parameters for this process:
\begin{description}
 \item[Max.edge] Maximum allowed triangle edge length, $1.5km$ within the study area and $7.5km$ in the boundary outside the area. 
 \item[Cutoff] Minimum allowed distance between points, $0.2km$.
 \item[Offset] Extension distance inside and outside the boundaries, $1.5km$ and $5km$, respectively. 
\end{description}
In addition to these parameters, the survey locations were also included in the construction of this mesh as a subset of the mesh node locations. 
This resulted in minimum edge length of $200m$ and maximum edge length of $1.5km$. The resulting model mesh is shown in Figure~\ref{fig:model_mesh}.

We implemented also BYM model using a triangular mesh. However, we used uniform mesh with constant edge length of $500m$, because we assume the resolution of the expert maps is such that the points within $250m$ of a node are indistinguishable from each other. We use a constant edge length because we want to weight each region equally in the likelihood - otherwise there would be more mesh nodes near the shoreline. 
Hence the effective resolution of the BYM model was coarser than that of the barrier model; the BYM mesh had approximately $2000$ mesh nodes whereas the barrier model had approximately $6000$ nodes. 
For the prediction raster we used cell size $150m \times 150$. These lower resolution rasters were construced with the R package raster. The covariate values and expert opinions of these new raster cells were calculated with the \verb!rasterize! function in the \verb!raster! package: The covariate values from the original raster were aggregated by taking their mean. For the expert opinion, the class with the highest probability among the aggregated cells was chosen.

After constructing the barrier model mesh, BYM-model mesh and the prediction raster we built the whole model on that by projecting the survey observations, expert assessments and the covariates onto the triangulation mesh. 
To do this, we first constructed the projection matrix $\mathbf{A}$ which projects function values from the triangulation mesh to the original 50m$\times$50m raster cells of environmental covariates.  That is,
\begin{equation}
\mathbf{X}_\text{raster} = \mathbf{A}\mathbf{X}_\text{mesh}
\end{equation}
In order for us to project from raster locations to the mesh, we take the matrix $\mathbf{A}$, transpose it, then normalize all rows that have non-zero elements to get transposed projection matrix $\tilde{\mathbf{A}}$. We then multiply the covariates and the expert assements from the original raster maps with $\tilde{\mathbf{A}}$ to get their projected values in the triangulation mesh:
\begin{align}
\mathbf{X}_\text{mesh} = \tilde{\mathbf{A}}\mathbf{X}_\text{raster} \\
\mathbf{Z}_\text{mesh} = \text{round}(\tilde{\mathbf{A}}\mathbf{Z}_\text{raster}) \nonumber
\end{align}
where $\mathbf{X}_\text{raster}$ is an $n\times 3$ matrix of covariates (one covariate per column) in the original raster data set and $\mathbf{Z}_\text{raster}$ is an $n\times 10$ matrix of expert assessments. Because expert assessments are categorical, we round the projected values.
The weights in $\mathbf{A}$ and $\tilde{\mathbf{A}}$ are based on the barycentric coordinates of each point $\textbf{s}$ projected onto the mesh. Each row of $\mathbf{A}$ has three positive elements summing to 1. These weights indicate how close the projected point is to the corresponding triangle vertex. 
In $\tilde{\mathbf{A}}$ the weights on rows determine how close corresponding raster points are to the given vertex. Some rows may have only zeroes on them, because the corresponding vertex is outside the model area.

\subsection{INLA stacks}
In the INLA model, the observation model is divided into one or multiple stacks that contain the observation data, prediction matrix $A$ and the effects used to predict the observations. Our model uses the following stacks:
\begin{description}
 \item[Survey] This stack contains the survey observations. The observation are predicted with the covariates $\textbf{x}$, intercept $\alpha$ and the survey random effect $\phi(\s)$.
 \item[Expert] This stack constain all the expert opinions. All opinions are included in a single vector, while the prediction matrix $A$ ensures that each element of this vector is predicted with the corresponding effects - the survey predictions with an expert intercept $\bar{\alpha}_j$, coefficient $\bar{c}_j$ and the expert random effect $\bar{\varphi}_j$.
 \item[Survey prediction] This stack is used to predict a replicate survey result for each cell of the prediction raster using the survey random effect as well as the fixed effects of the covariates. The stack has similar structure to the survey stack, the difference is that now all observation are missing values, i.e. they have to be predicted and the predicted points are now the raster cells instead of the survey locations.
 \item[Expert prediction] This stack is used to predict the expert opinion of each expert for each cell of the prediction raster.
 \item[Zero stack] This is a special stack used to connect the survey predictions to the expert stacks. This is done, because with INLA we cannot directly use the same coefficient for multiple effects, so we instead create a fixed effect that has the same distribution as the combination of these effects. \cite{ruiz2012direct}
\end{description}
In all of the stacks, the prediction matrix $A$ maps the effects to the corresponding observations. This is needed, because the random effects are calculated along the mesh nodes, while the observations and predictions are either on the survey locations or the raster cells. Additionally, some effects only apply to a subset of the observations: the expert opinions only apply to certain raster cells. The prediction matrix is constructed using the INLA function \verb!inla.spde.make.A!.                                                                                                                                                                                                             
For more details on implementation see the electronic supplement that contains all the data and code used to conduct this study.

\section{Likelihood functions}\label{sec:Appendix_for_likelihoods}

\begin{figure}[t]
\begin{subfigure}[b]{\textwidth}
	\hspace{0cm}\includegraphics[scale=0.65]{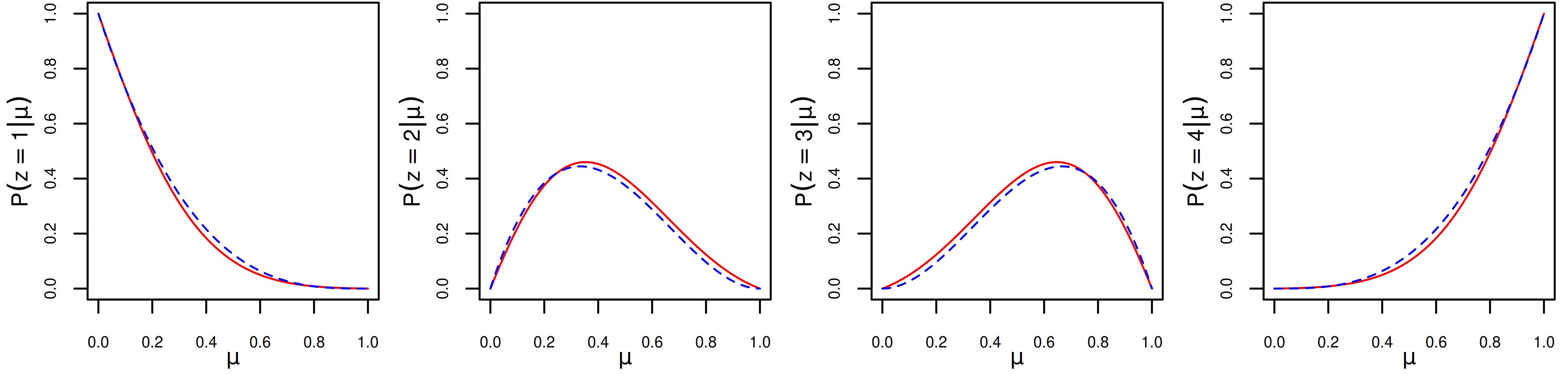}
\end{subfigure}
\begin{subfigure}[b]{\textwidth}
	\hspace{0cm}\includegraphics[scale=0.65]{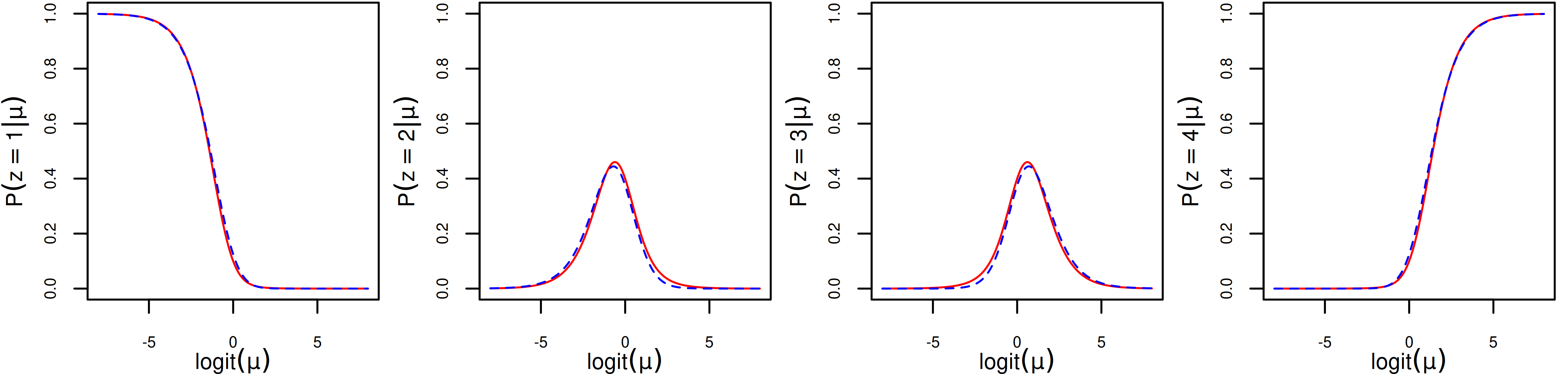}
\end{subfigure}
\caption{The exact and the approximate (dashed line) likelihood functions.}\label{fig:likelihood_functions}
\end{figure}

We approximated the model for experts' presence-absence assessments (3.8) with logit-Bernoulli model

\begin{equation}\label{eq:expert_assessment_jointModel}
\mathbbm{1}(z_{ij}\in \{3,4\})|\bar{\pi}_{ji} \sim \mathrm{Bernoulli}(\bar{\pi}_{ji}),
\end{equation}
where $\bar{\pi}_{ji}$ is given in (3.6).
When using all categories of expert assessments, we approximate the likelihood functions arizing from Beta cumulative distribution functions with the Binomial probability. We chose the corresponding binomial probability parameters $N$ and $\psi$ for each expert assessment category $z\in\{1,\dots,4\}$ by finding the least squares fit to the true likelihood function (3.9). That is we solved for
\begin{align}
\hat{N}_z,\hat{\psi}_z &= \argmin_{N,\psi} \int_0^1 (\text{Pr}(z|\mu ,s_j)-\text{Bin}(\psi|N,\mu))^2 d\mu \\
& \approx \argmin_{N,\psi} \frac{1}{n} \sum_{i=1}^n (\text{Pr}(z|\frac{i}{n},s_j)-\text{Bin}(\psi|N,\frac{i}{n}))^2\label{eq:beta_cdf_ssq}
\end{align}
where $s_j=2$. 
We found the optimal fit through grid search, i.e. calculating \eqref{eq:beta_cdf_ssq} for a range of values of $N$ and $\psi$. We found that with $\hat{N}_z=3$ for all $z\in\{1,\dots,4\}$ and $\hat{\psi}_1=0, \hat{\psi}_2=1, \hat{\psi}_3=2$ and $\hat{\psi}_4=3$ the Binomial approximations were closest to the true likelihoods. The match between the approximation and the exact likelihood functions is visualized in Figure~\ref{fig:likelihood_functions}.

\bibliographystyle{ba}


\section{Extra results}\label{sec:Extra_results}

In this section we collect additional result figures.

\begin{figure}[t]
\hspace{-0.5cm}\begin{subfigure}[b]{0.31\textwidth}
	\includegraphics[scale=0.55]{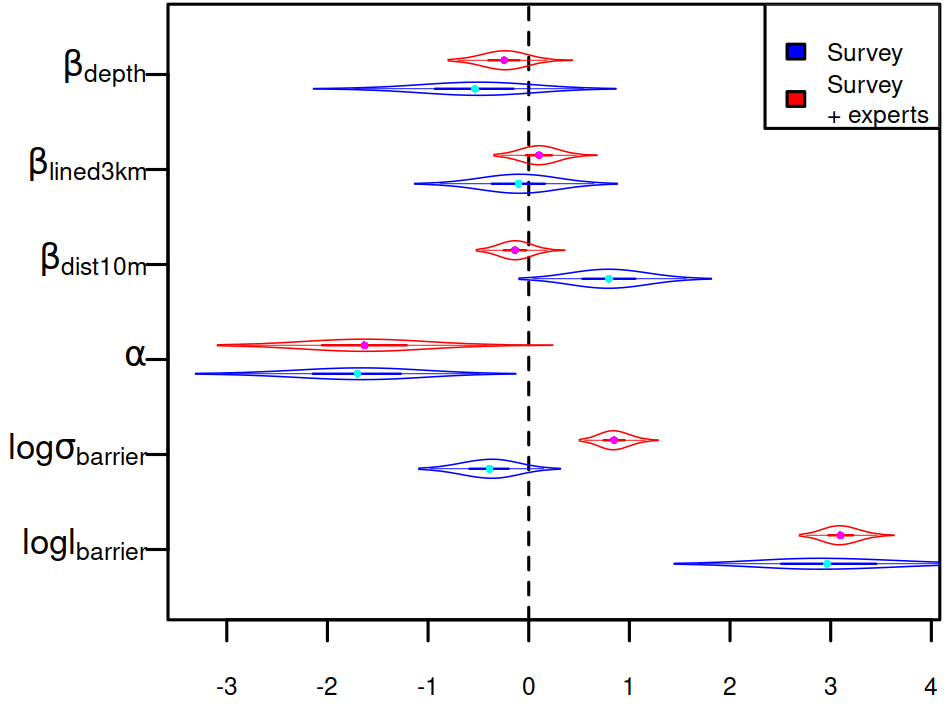}
	\caption{survey=p/a, expert=p/a}
\end{subfigure}
\hspace{0.4cm}\begin{subfigure}[b]{0.31\textwidth}
\includegraphics[scale=0.55]{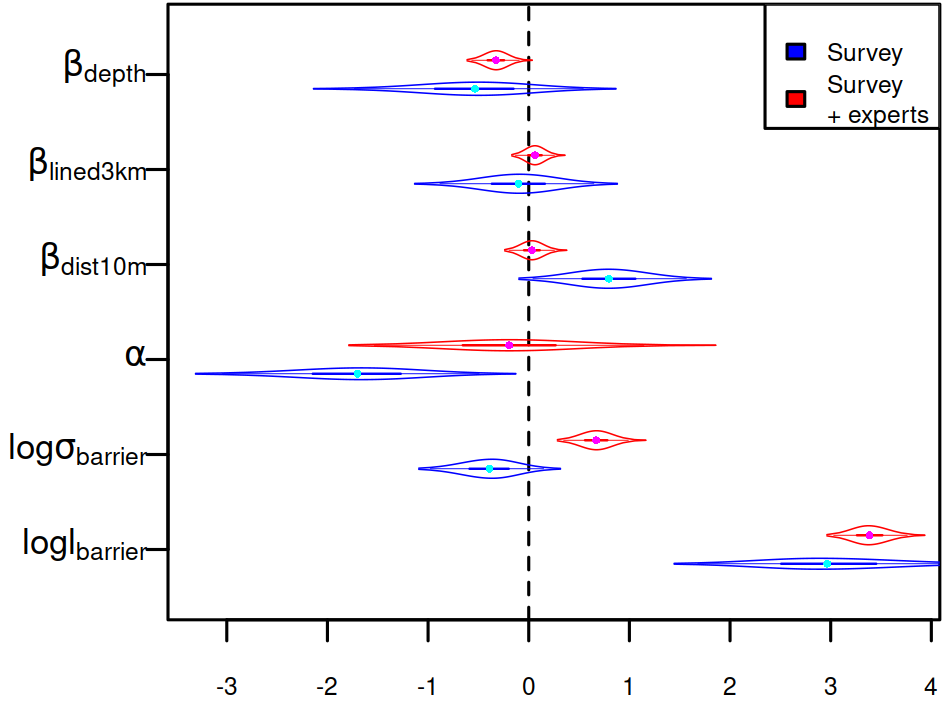}
	\caption{survey=p/a, expert=abu}
\end{subfigure}
\hspace{0.4cm}\begin{subfigure}[b]{0.31\textwidth}
	\includegraphics[scale=0.55]{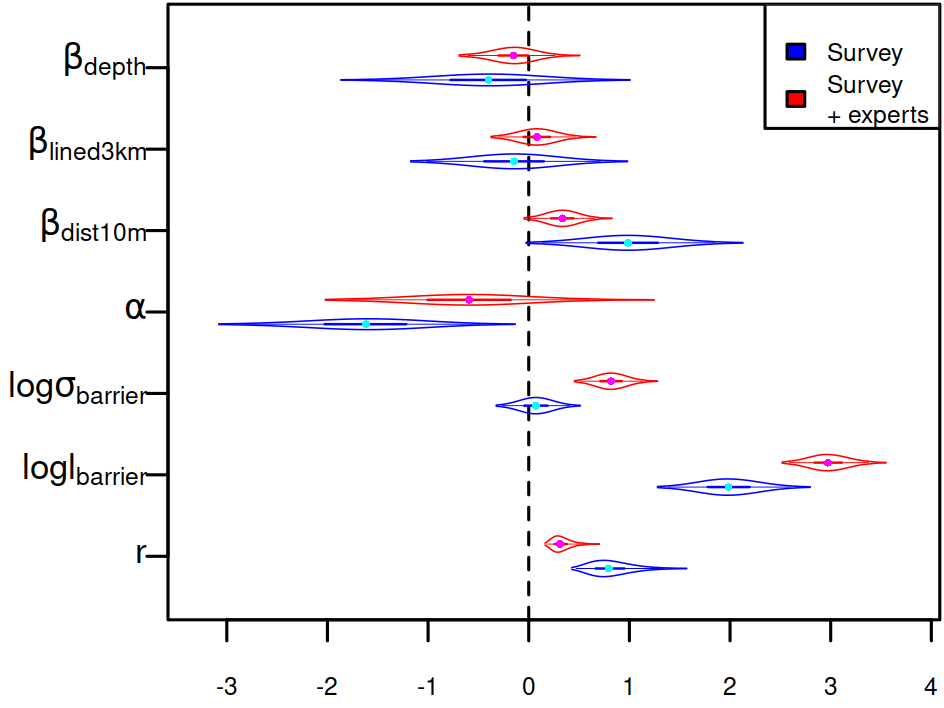}
	\caption{survey=abu, expert=p/a}
\end{subfigure}
\caption{Survey model parameters in alternative models; "p/a" denotes the presence absence model ((3.4) for survey data and (3.8) for expert assessments), and "abu" denotes abundance model (3.3) for survey data and four categories model (3.9) for expert assessments.}\label{fig:survey_parameters_extra}
\end{figure}

\begin{figure}[t]
\hspace{-0.5cm}\begin{subfigure}[b]{0.31\textwidth}
	\includegraphics[scale=0.55]{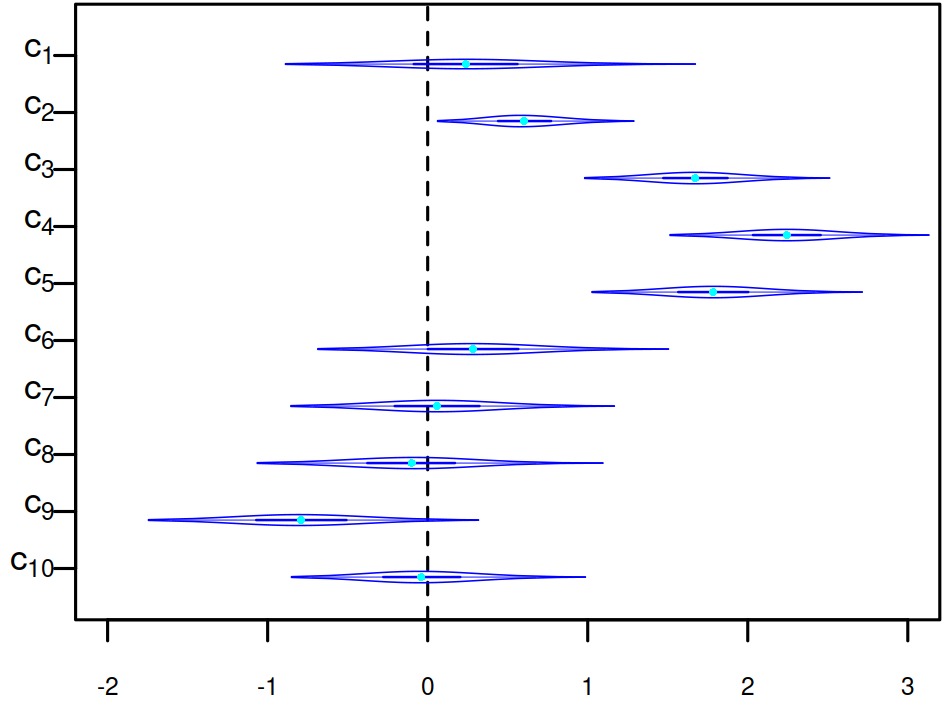}
	\caption{survey=p/a, expert=p/a}
\end{subfigure}
\hspace{0.4cm}\begin{subfigure}[b]{0.31\textwidth}
\includegraphics[scale=0.55]{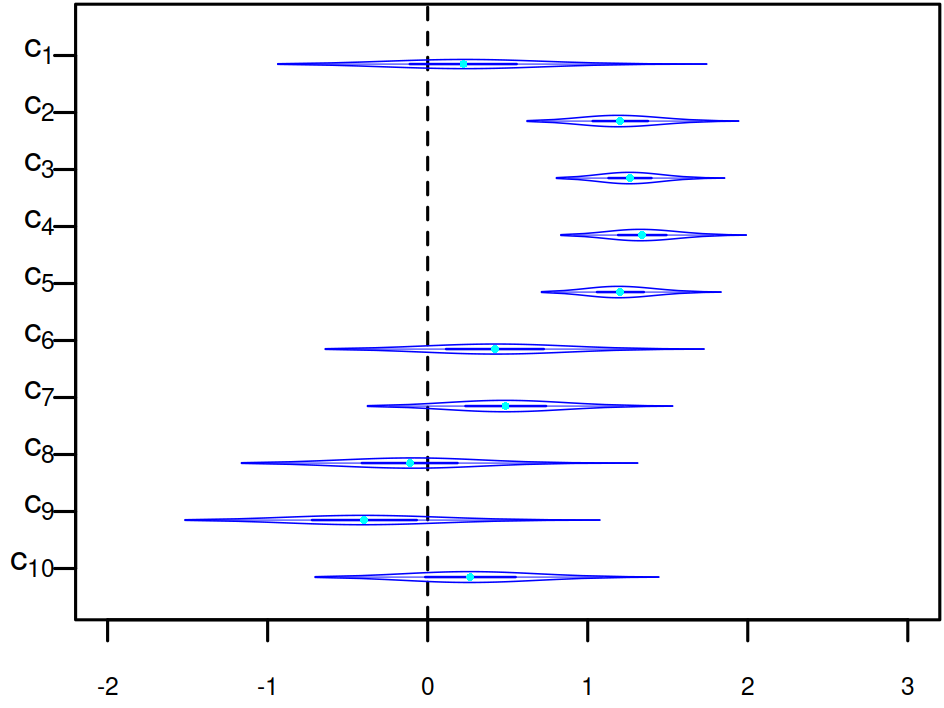}
	\caption{survey=p/a, expert=abu}
\end{subfigure}
\hspace{0.4cm}\begin{subfigure}[b]{0.31\textwidth}
	\includegraphics[scale=0.55]{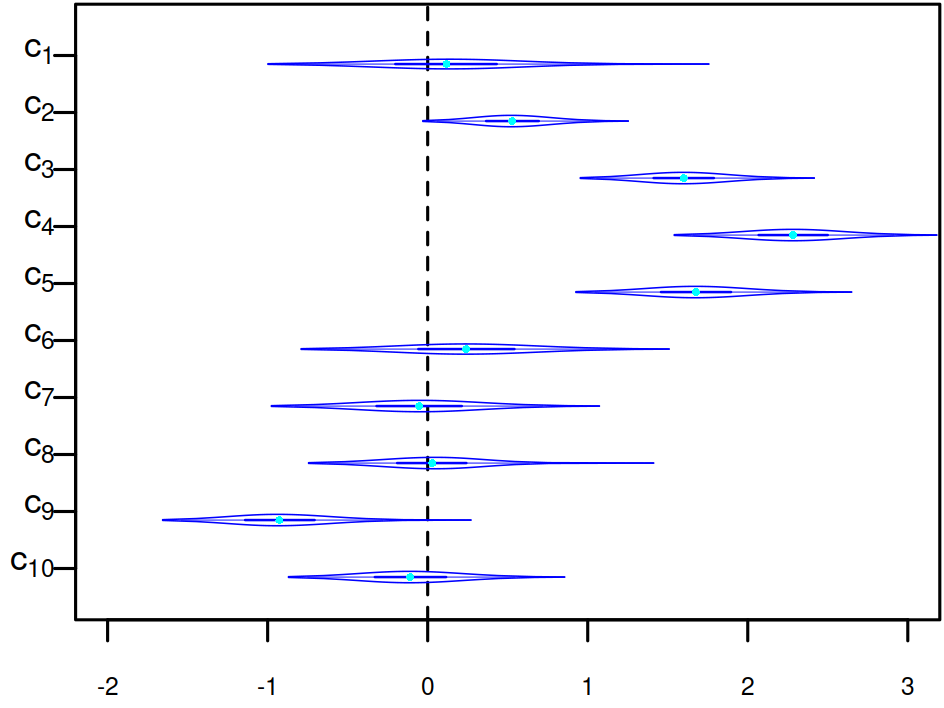}
	\caption{survey=abu, expert=p/a}
\end{subfigure}
\caption{Expert coefficients in alternative models; "p/a" denotes the presence absence model ((3.4) for survey data and (3.8) for expert assessments), and "abu" denotes abundance model (3.3) for survey data and four categories model (3.9) for expert assessments.}\label{fig:expert_coefficients_extra}
\end{figure}

\begin{sidewaysfigure}
    \centering
    \includegraphics[scale=0.65]{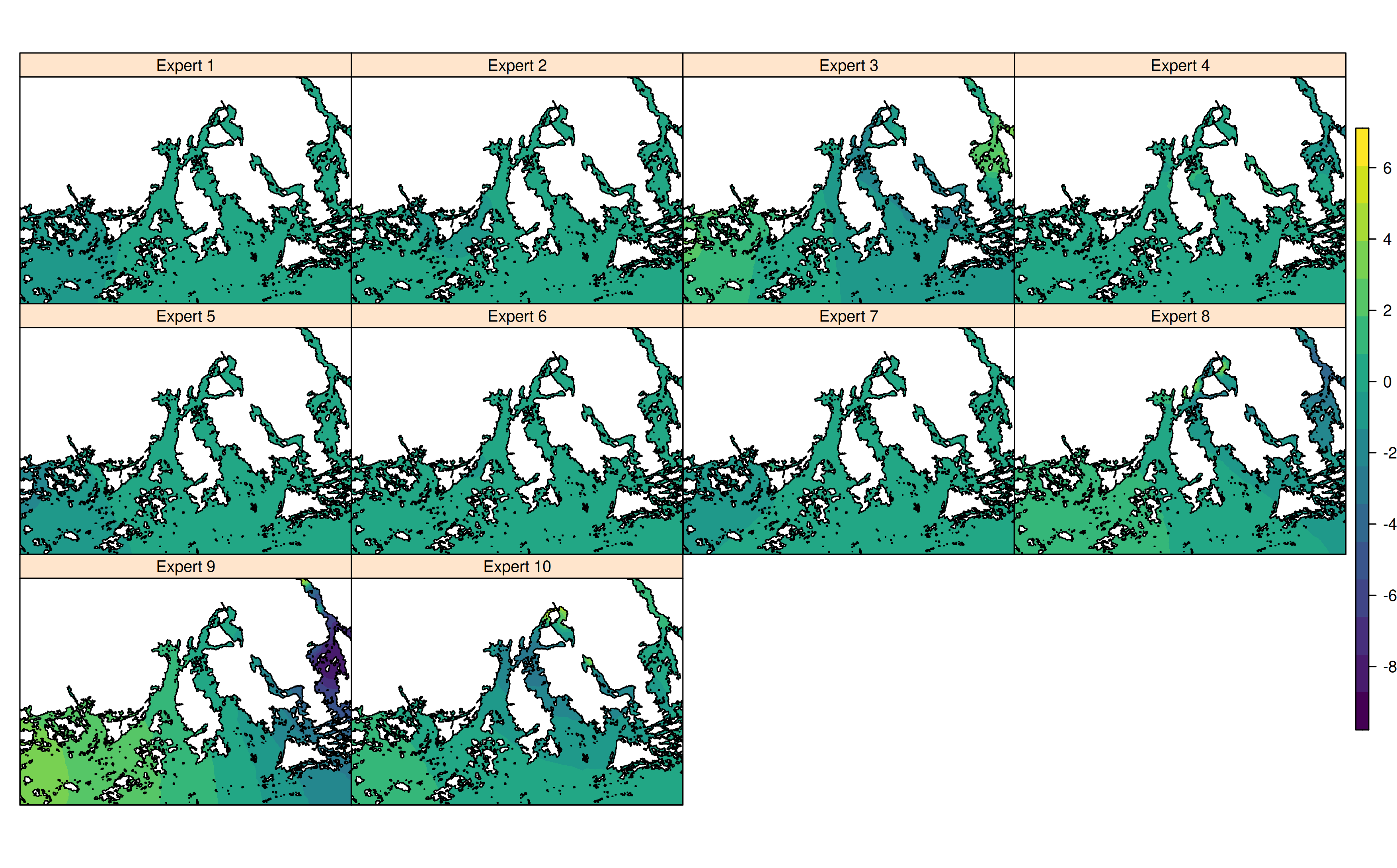}
    \caption{The posterior mean of the spatial bias terms, $\bar{\varphi}$, for the experts.}\label{fig:bym_mu}
\end{sidewaysfigure}

\begin{figure}
\begin{subfigure}[b]{\textwidth}
	\hspace{-.6cm}\includegraphics[scale=0.75]{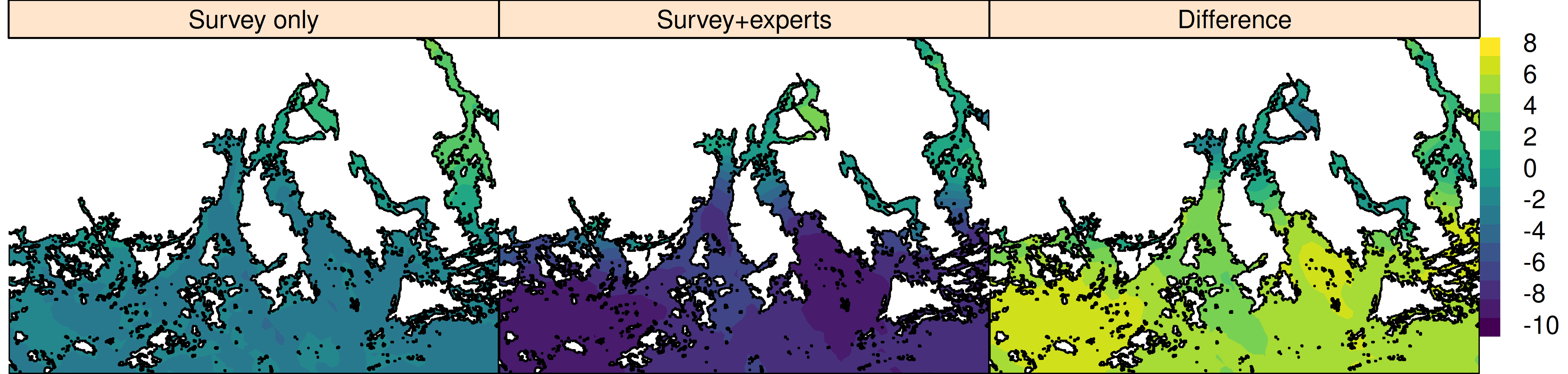}
\end{subfigure}
\begin{subfigure}[b]{\textwidth}
	\hspace{-.6cm}\includegraphics[scale=0.745]{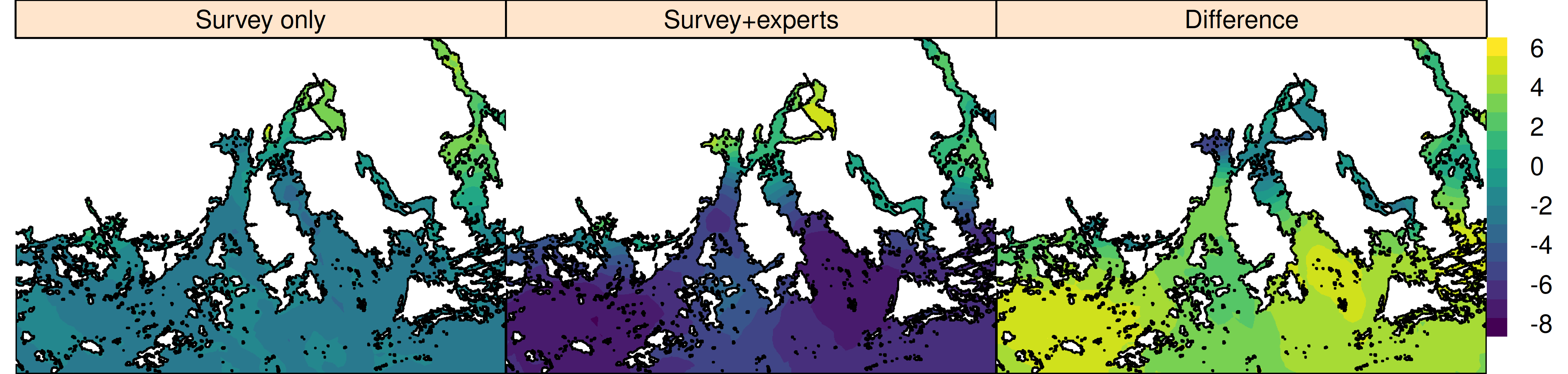}
\caption{Posterior predictive mean.}\label{fig:predictive_density_bernoulliExpert}
\end{subfigure}
\begin{subfigure}[b]{\textwidth}
	\hspace{-.6cm}\includegraphics[scale=0.743]{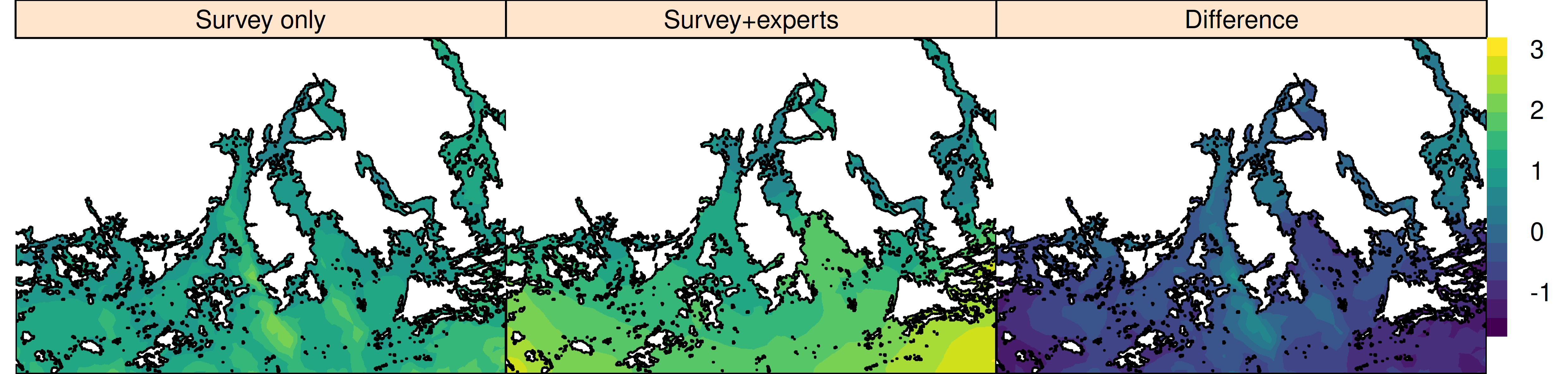}
\end{subfigure}
\begin{subfigure}[b]{\textwidth}
	\hspace{-.6cm}\includegraphics[scale=0.75]{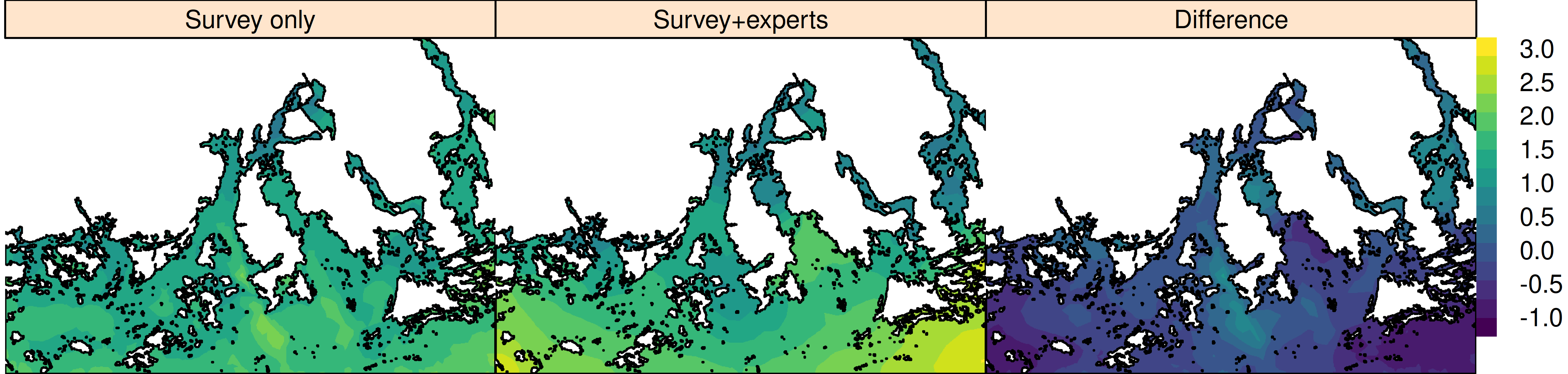}
	\caption{Posterior predictive standard deviation.}\label{fig:predictive_density_bernoulliExpert_sd}
\end{subfigure}
\caption{Posterior predictions (a) and their uncertainty (b) for larval distribution in the study area with survey only and survey+experts models and their difference ([survey only] - [survey+experts]). In both subplots, the top row shows results when the survey data are modeled as occurrence data and the maps show the posterior predictive mean of log odds ratio (a) or its standard deviation (b) for larval occurrence and their difference. In both subplots, the bottom row shows results when the survey data are modeled as count data and the maps show the posterior predictive mean of log larval density (a) or its standard deviation (b) and their difference. The expert assessments are modeled with Bernoulli observation model in both cases.}
\end{figure}

\bibliography{references}